\documentclass[pre,reprint,letterpaper,numerical]{revtex4-1}
\usepackage{color}
\usepackage{geometry}
\usepackage{graphicx}
\usepackage{amssymb}
\usepackage{bm}
\usepackage{amsmath}

\usepackage{ulem}

\begin{document}

\title{How hydrophobic drying forces impact the kinetics of molecular recognition}

\author{Jagannath Mondal} 
\affiliation{Department of Chemistry, Columbia University, New York, NY 10027}

\author{Joseph A. Morrone}
\affiliation{Department of Chemistry, Columbia University, New York, NY 10027}

\author{B. J. Berne}
\email{bb8@columbia.edu}
\affiliation{Department of Chemistry, Columbia University, New York, NY 10027}

\begin{abstract} 
  A model of protein-ligand binding kinetics in which slow solvent
  dynamics results from hydrophobic drying transitions is
  investigated. Molecular dynamics simulations show that solvent in
  the receptor pocket can fluctuate between wet and dry states with
  lifetimes in each state that are long enough for the extraction of a
  separable potential of mean force and wet-to-dry transitions.  We
  introduce a Diffusive Surface Hopping Model that is represented by a
  two-dimensional Markovian master equation. One dimension is the
  standard reaction coordinate, the ligand-pocket separation, and the
  other is the solvent state in the region between ligand and binding
  pocket which specifies whether it is wet or dry.  In our model, the
  ligand diffuses on a dynamic free energy surface which undergoes
  kinetic transitions between the wet and dry states. The model yields
  good agreement with results from explicit solvent molecular dynamics
  simulation and an improved description of the kinetics of
  hydrophobic assembly.  Furthermore, it is consistent with a
  ``non-Markovian Brownian theory'' for the ligand-pocket separation
  coordinate alone.
\end{abstract}

\maketitle

\section{Introduction}
\label{sec:Intro}

Recent theoretical work has shown that the displacement of
  water by drug molecules is important in the thermodynamics
  and kinetics of ligand--enzyme binding~{\cite{young07,shan11,dror2011}}.
  The kinetics of drug docking are known to be a key metric for lead
  optimization~\cite{copeland06}. In the present work, we explore the
  kinetic motifs of hydrophobic association on ligand binding. This is
  achieved by the development of a simple model for
  hydrophobic association that is compared with explicit solvent
  molecular dynamics simulations.

One of the signature features of hydrophobic assembly is the
observation of a dewetting
transition~{\cite{berne_rev,Hummer_rev,chandler_rev, sharma12,sharma12-2}}.  Drying
behavior has been found to play an important role in protein
self-assembly and the behavior of nano-confined
water~\cite{liu05,Hummer_cnt}. The present paper draws on our
extensive work on the role of molecular-scale hydrodynamics in
hydrophobic collapse~\cite{joe12,jingyuan12} where we showed that when
the attraction between water and two associating nanoscale objects is
weak, assembly proceeds via a drying transition in the inter-solute
region. This transition is characterized by large peaks in the
relative translational friction coefficient that correspond to large
and slow solvent fluctuations.  The slow relaxation times exhibited by
water undergoing dewetting transitions suggest that resulting
non-Markovian effects may prove to be a crucial element in a full
description of the assembly kinetics.

We presently extend our earlier investigations to a model of a
spherical ligand docking in a concave cavity.  The model is similar to
one investigated in a series of papers by McCammon and
co-workers~\cite{Baron:JACS10,baron13}, but altered to describe the
assembly of a nanoscale ligand. This alteration facilitates the study
of a large-scale drying transition.  We investigate the
molecular-scale hydrodynamic effects and the kinetic rate constants
for binding and develop a theoretical framework to describe
hydrophobic assembly which couples the diffusive reaction coordinate
(the separation) to transitions between ``wet'' and ``dry'' states
which are defined by a coarse-grained solvent occupancy in the binding
pocket. This model is conceptually similar to the surface hopping
algorithm employed in non-adiabatic quantum dynamics \cite{tully12},
and thus we call it the diffusive surface hopping model (DSHM). We
show how the model reproduces the effect of drying fluctuations that
are evidenced in an ensemble of explicit solvent molecular dynamics
assembly trajectories. We compare different kinetic schemes with the
results of all-atom molecular dynamics.

In very recent work, Setny, {\it et al.}~\cite{setny13} have computed
the hydrodynamic profile for the ligand binding model that originated
in Ref. \onlinecite{Baron:JACS10}.  It was found that enhanced and slowed
hydration fluctuations engender a slow down in the ligand dynamics, in
agreement with our results on model plates and
spheres~\cite{joe12,jingyuan12}.  This work also reported a 
shift in the spatial  hydrodynamic effect that was attributed to non-Markovian effects.  We find a related behavior in our study
and show that it is resolved by use of the DSHM.  In this way, the
theoretical framework that is presently introduced yields a
coarse-grained dynamical scheme that improves upon the description
obtained from Smoluchowski (Brownian) dynamics when slow solvent
motions are important.

\section{Coarse grained descriptions of hydrophobic assembly}
Ligand binding kinetics is often described by the Smoluchowski
Equation: 
\begin{eqnarray}
  \frac{\partial p(q,t)}{\partial t} = \frac{\partial}{\partial q} D(q) \left( \frac{\partial}{\partial q} - \beta\bar{F}(q) \right) p(q,t) 
\label{eq:smoul}
\end{eqnarray}
where $q$ is the separation between ligand and receptor, $p(q,t)$ 
is the time-dependent probability distribution function,
$\bar{F}(q)=-\partial W(q)/\partial q$ is the mean force, $W(q)$ is
the potential of mean force, $D(q)=k_bT/\zeta(q)$ is the spatially
dependent diffusion coefficient, and $\zeta(q)$ is the friction
coefficient. This is a Markovian equation and is valid if solvent
fluctuations are very fast compared to solute motions. The spatial
dependence of the diffusion coefficient arises from hydrodynamic
interactions (HI) between the receptor and ligand.  $\bar{F}(q)$ and
the $\zeta(q)$ may be computed from MD~\cite{joe12,jingyuan12}.  Eq.
(\ref{eq:smoul}) was tested in our previous work but we observed very
slow solvent fluctuations at and around the drying transition between
hydrophobic plates (or spheres). Indeed the autocorrelation function
of the solvent force along $q$ exhibited  prominent long time tails,
indicating that non-Markovian effects should be important.  In such
cases solvent degrees of freedom must be included in describing
hydrophobic assembly\cite{liu05,willard08}. We now develop a theory
that is applicable to the ligand-receptor model. It includes a coarse
grained description of the solvent as an explicit degree of freedom of
the dynamics, and involves a two dimensional Smoluchowski equation,
which although Markovian, yields a non-Markovian expression for
$p(q,t)$ in place of Eq. (\ref{eq:smoul}), when the solvent degree of
freedom is projected out.

\begin{figure*}[t]
\begin{center}
\includegraphics[scale=0.50]{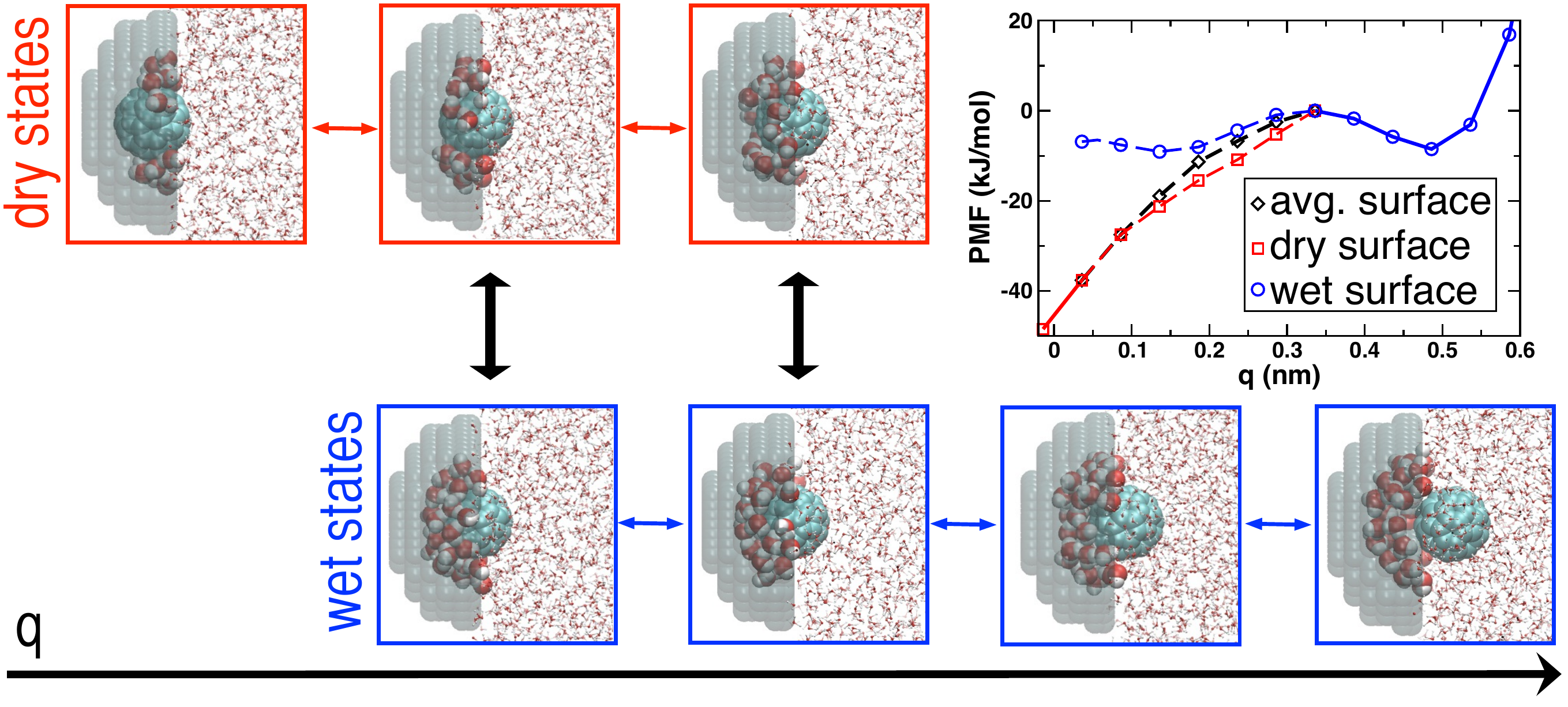}
\end{center}
\caption{Schematic for the Markovian master equation analysis that is
  presently employed.  Transitions amongst wet states (blue framed
  snapshots) and dry states (red framed snapshots) occur along $q$.
  Transitions between wet and dry states may occur in the dewetting
  region at fixed $q$.  In the upper-right panel the potential of mean
  force is plotted on the wet surface (blue line) and dry surface (red
  line).  In the region of drying, two of the models presently
  considered evolve along an averaged surface (black dashed line).
  Snapshots are rendered with VMD~\cite{vmd96}. }
\label{fig:mastereq}
\end{figure*}

We propose a two dimensional model where one coordinate is the
diffusive coordinate $q$ (the separation between receptor and ligand),
and the other is a discrete state variable $s=w$ or $d$,
indicating whether the binding pocket is wet or dry, respectively. 
This model has state dependent diffusion coefficients, $D(w,q)$ and $D(d,q)$,
evolves on state-dependent free energy surfaces, $W(w,q)$ and 
$W(d,q)$, 
and its state can change from wet to dry and from dry to wet
by first order kinetics with rate constants also depending on $q$,
namely $k(d\leftarrow w, q)$ and $k(w\leftarrow d, q)$ are
 the rate constants to transition from $w\rightarrow d$ and $d
\rightarrow w$, respectively.  This model is equivalent to a particle diffusing on a
potential energy surface that can hop between two functional forms,
one for the wet and one for dry states, with different spatially
dependent diffusion constants on each surface. We call this the
Diffusive Surface Hopping Model (DSHM).  Such
a scheme can be described by the following two differential equation
for the time evolution of the probability density $p(s,q,t)$:
\begin{eqnarray} \frac{\partial p(s,q,t)}{\partial t} =
\frac{\partial}{\partial q} D(s,q) \left( \frac{\partial}{\partial q}
- \beta\bar{F}(s,q) \right) p(s,q,t) \nonumber \\ - k(s^\prime
\leftarrow s,q) p(s,q,t) + k(s \leftarrow s^\prime,q)p(s^\prime,q,t)
\; , \nonumber \\ \label{eq:motion_eq}
\end{eqnarray} 
where $\bar{F}(s,q) = -\partial_q W(s,q)$.  One equation where $s=w$ and $s^\prime = d$ is paired to one 
corresponding to $s=d$ and $s^\prime = w$.  
In this way, the diffusion dynamics is coupled to transitions between the surfaces.

DSHM reduces to Eq. (\ref{eq:smoul}) when the hopping rate between
surfaces is fast compared to the rate of diffusion along $q$ (see
Appendix).  This model is similar in spirit to one
discussed in Ref. \onlinecite{bernebook} where diffusing charged particles in
an electric field can hop between two states with different diffusion
coefficients and electric mobilities, with the hopping governed by
first order chemical kinetics. However in this prior formulation, a
treatment of the spatial dependence of the diffusion coefficients,
electrical mobilities, and transition rates is ignored.

Application of Eq. (\ref{eq:motion_eq}) calls for specifying mean force
and diffusion coefficient separately for both the wet and dry states
as well as a set of transition rates between these states. The problem
can be simplified for our receptor-ligand problem since transitions
between wet and dry states only take place for separations in a narrow
range (in the neighborhood of the critical distance for drying
$q_c$). We assume that for large values of $q$ only wet states are
accessible and for small values of $q$ only dry states are accessible.
Then one need only consider transitions between surfaces in a specific
``drying region.''  One can then discretize the continuous
Smoluchowski dynamics~\cite{szabo98}, and place both Eq.
(\ref{eq:smoul}) and Eq. (\ref{eq:motion_eq}) in the form of a
Markovian Master Equation:
\begin{equation} \frac{\partial p_i(t)}{\partial t} = R_{ij} p_j(t) \; , 
\label{eq:mastereq}
\end{equation}
where the index i runs over all allowed states, and where q is
represented on a grid.  Eq. (\ref{eq:motion_eq})
describes evolution of two surfaces, both wet and dry.  The detailed
expressions for $R_{ij}$ in the case of both one dimensional
Smoluchowski (Brownian) dynamics and our two dimensional two surface
representation are given in the Appendix, and a
schematic depicting how transitions are made in two surface model is
depicted in Fig.~\ref{fig:mastereq}.  Markovian Master Equations
(Markov State models) can serve as an important tool to analyze
conformational changes in biomolecules and extract kinetic information
from molecular simulation~\cite{pande2010}, and have been also
utilized to treat solvent degrees of freedom, including drying
fluctuations in carbon nanotubes~\cite{sriraman05,gu2013}.

The elements of the rate matrix $\bm{R}$ given in
Eq. (\ref{eq:mastereq}) are obtained from a molecular dynamics
simulation runs in explicit solvent where the ligand-pocket separation
$q$ is restrained to a set of values along the assembly path.  The
model ligand is a C60 fullerene and the pocket is an ellipsoidal hole
carved from a hydrophobic wall. Further details are given in the Appendix.

\section{Results and Discussion}
\subsection{Average mean force and hydrodynamic profiles} 
The nature of the free energy surface and hydrodynamic interactions
that the solute experiences depends intrinsically on the strength of
the solute-water interaction.  Vastly different behavior is exhibited
in the case of very hydrophobic bodies where assembly is facilitated
by drying as compared with more hydrophilic bodies where steric
interactions engender the expulsion of water at small separations.  We
have computed the potential of mean force and hydrodynamic profile for
the model pocket for three different strengths of solvent attraction.
The interactions that describe the ligand are not varied.  The weakest
and intermediate interactions conform to the case of hydrophobic
assembly driven by drying transitions, whereas the behavior of the
 strongly attractive pocket is dominated by steric ejections of water. 
 The intermediate strength of attraction will be the focus
of this work and discussion of the other two cases is presented in the
Appendix.

\begin{figure}[t]
\begin{center}
\includegraphics[width=3.0in]{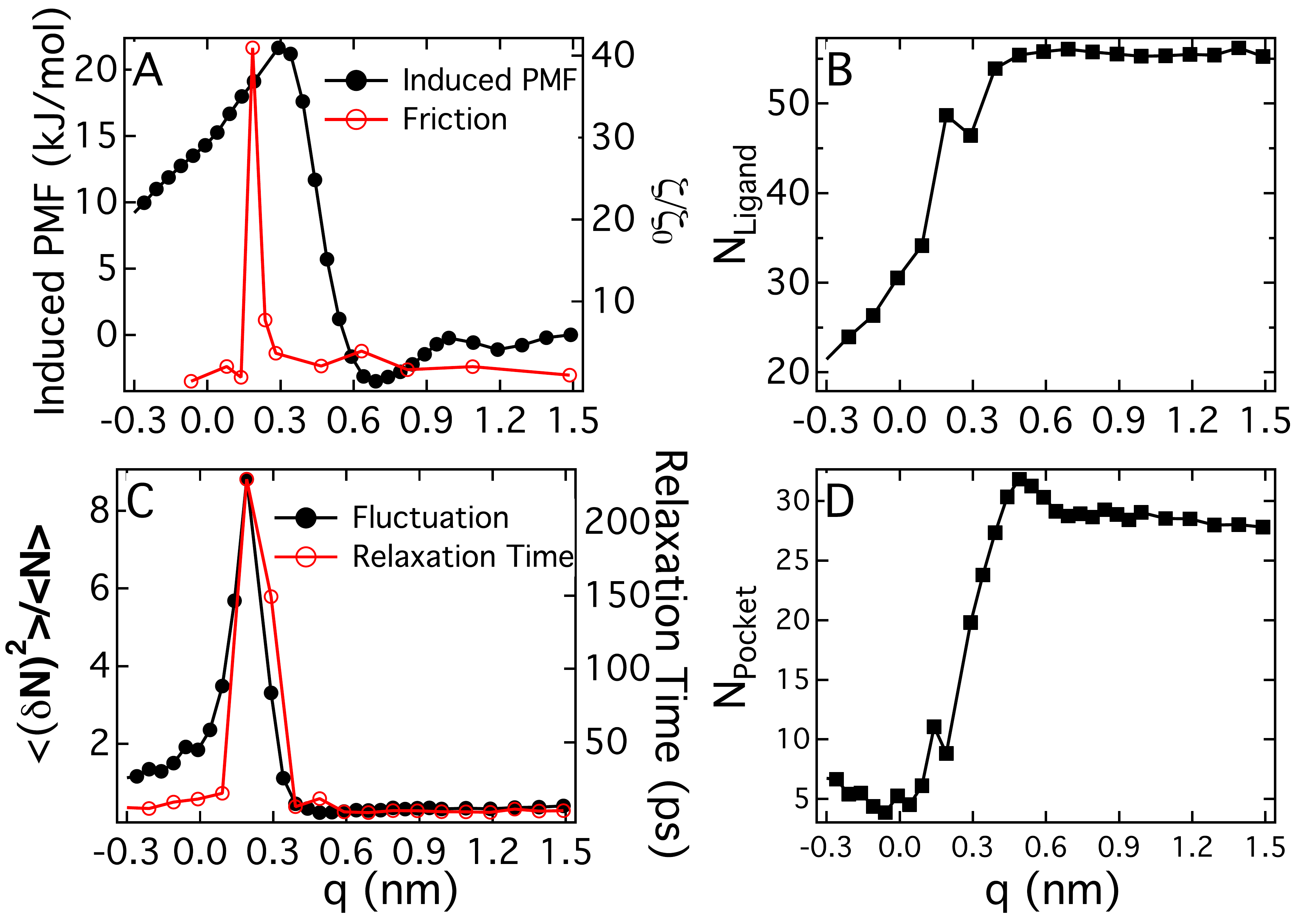}
\end{center}
\caption{Layout describing the correlations among various
  thermodynamic and hydrodynamic profiles as a function of
  ligand-pocket distance for intermediate-attractive pocket. (A) Comparison of
  the solvent-induced PMF (left scale) and friction profile(right scale), (B) 
  ligand-water number profile, (C) Correlation of
  pocket-water fluctuation (left scale) and relaxation time (right scale), and (D) pocket-water number profile.}
\label{fig:Layout_semi}
\end{figure}

Figs.~\ref{fig:Layout_semi}B and \ref{fig:Layout_semi}D
 show how the number of water molecules in the first
  solvation shell of the ligand, $N_\text{Ligand}$ and the number of pocket
  water molecules $N_\text{Pocket}$ vary as a function of the reaction
  coordinate $q$ for intermediate pocket-water attraction.  As the
ligand enters the pocket, the system undergoes a drying transition
around $q_c = 0.186$ nm. As the ligand is rather hydrophilic, there is a
free energy penalty associated with the stripping away of waters from
the fullerene.  There is also an observed maximum in the number of
pocket water molecules as the ligand approaches the cavity, owing to
the intrusion of the ligand solvation shell into the pocket.

The solvent-induced potential of mean force and variation of the
friction coefficient with separation $q$ are plotted in
Fig.~\ref{fig:Layout_semi}A.  Unlike our previous work where the
friction coefficient was computed from the force-force autocorrelation
function,~\cite{Bocquet:1997p103}, we presently utilize a technique
that applies a harmonic restraint along the $q$-direction at several
positions and probes the relaxation of the position autocorrelation
function~\cite{Straub:1990p65,hummer05}.  The region of drying is
associated with a large peak in the friction coefficient, $\zeta(q)$.
The mean solvent induced force, is attractive near $q_c$. For the
present system, barriers primarily arise from the desolvation of the
ligand (see Fig.~\ref{fig:Layout_semi}B).

Plotted in Fig.~\ref{fig:Layout_semi}C are the solvent
fluctuations and relaxation times in the binding pocket as a function
of $q$.  These properties have been found to yield trends relatable to
those observed for the profile of the friction
coefficient in our previous works~\cite{joe12,jingyuan12}, and this is also observed in the
present work.  It is seen that relaxation times greater than $200$ ps
are present in the cavity near the dewetting transition.  As we will
show later, this timescale is of the same order as the mean first
passage time for the ligand to bind to the pocket starting from entry
into the pocket.  Such slow fluctuations indicate that
the simple one dimensional Brownian dynamics approximation does not
hold, and non-Markovian effects are significant.

\subsection{Hydrophobic forces on wet and dry surfaces} 
\begin{figure}[t]
\begin{center}
\includegraphics[width=3.0in]{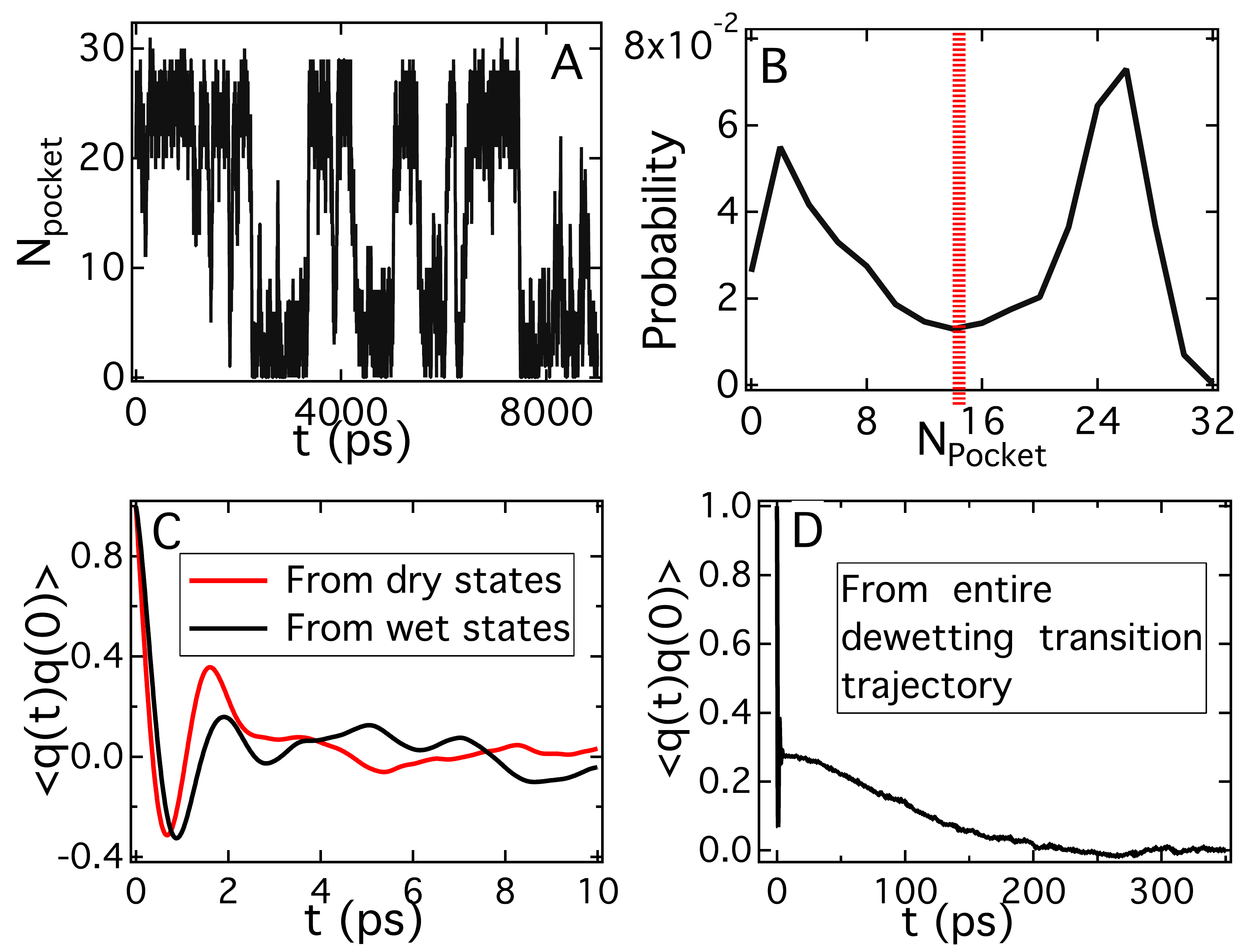}
\end{center}
\caption{(A) Representative trajectory exhibiting continuous
  fluctuations between wet and dry state at 
  $q_c=0.186$ nm. (B) The probability
  distribution of solvent occupancy in the pocket at $q_c$.
  Configurations to the left and right of the dashed red line are
  considered to be in ``dry'' or ``wet'' states, respectively.  The normalized
  position autocorrelation function about
  $\left<q\right>=q_c$ is plotted in the wet (black line) and dry (red
  line) states in panel (C) and the total autocorrelation function
  from which the friction coefficient at $q_c$ is determined is
  plotted in panel (D). The long tail of
  the total correlation function is absent in the wet and dry state
  results, indicating that the large friction is associated with
  slow solvent fluctuations due to drying and wetting events.}
\label{fig:wetdry}
\end{figure}

A detailed characterization of the underlying solvent coordinate in
the DSHM calls for a quantitative analysis of the dry and wet states
observed in the dewetting region. The probability distribution and
representative trajectory of the pocket water molecules are shown for
$q= q_c$ in Fig.~\ref{fig:wetdry} (A) and (B). From the plot of the
number of waters in the region between ligand and pocket versus time
in panel A, one sees that the wet and dry states have sufficiently
long survival times, and therefore various average properties for the
wet and dry states can be determined.

To underline the importance of the slow fluctuations between wet and
dry states at the dewetting transition, we plot the normalized
  position autocorrelation function of the total system in
  Fig.~\ref{fig:wetdry}D and for the wet and dry states separately
  in Fig.~\ref{fig:wetdry}C. One can see that the relaxation
times of the correlation functions are markedly shorter when the
states are considered separately.  The friction coefficient in the wet
and dry states may be estimated from the correlation function $\langle
q(0)q(t)\rangle$ using the approach in
Refs. \cite{Straub:1990p65,hummer05}.  The friction in the wet state
is close to the value found at large separations, whereas in the dry
state it is approximately half of this. Therefore the pronounced
hydrodynamic effect we observe at the drying transition in this and
prior studies is shown to be due to slow transitions between the wet
and dry states, and it is appealing to incorporate this as a separate
slow collective variable, whereas other solvent degrees of freedom
remain treatable in the Markovian limit. Indeed, this is the rationale
behind the Diffusive Surface Hopping Model.

\subsection{Constructing the Diffusive Surface Hopping Model} 
The parameterization of our Diffusive Surface Hopping Model (DSHM)
draws from the underlying simulation results as obtained in the region
of the drying transition. As evident from Eq. (\ref{eq:motion_eq}), the
three main inputs necessary for DSHM are the mean force and diffusion
coefficient along the two surfaces and the rate constants for
transitions between wet and dry states.  Here we will briefly outline
only the salient features of how we construct the DSHM using
simulation as the source of parameters. A detailed discussion of the
parameterization is reserved for the Appendix.

The DSHM requires mean forces on the dry and wet surfaces. It consists
of numerous wet and dry states set on a equally spaced grid along
$q$. For the purpose of this model, a state is defined as dry if there
are fewer than 15 water molecules in the pocket, and wet otherwise. We
extract mean ``dry'' and ``wet'' forces by averaging over wet and dry
configurations separately at each fixed $q$ in the dewetting
region. The resulting potentials of mean forces, which include the
direct interactions of heavy bodies, that correspond to the wet and
dry surfaces are plotted in Fig. \ref{fig:mastereq}.

Another crucial set of input parameters for DSHM are the rate
constants for the transitions,
$\mbox{wet}\rightleftharpoons\mbox{dry}$. Such transitions are only
treated in the dewetting region.  Only wet and dry states are
considered at large and small separations, respectively.  The matrix
elements are estimated from the average dwell times in the wet state
from MD simulations at fixed $q$.  The reverse transitions are then
estimated from the detailed balance condition and the equilibrium
probabilities of being in a wet or dry state. The values of the
inverse rate constants at the values of $q$ considered are given in
Table 1.  The transition times are shown to become shorter as the
bodies approach each other and fewer water molecules are displaced by
the drying event.  Recent work on the rate of
drying~\cite{sharma12,sharma12-2} finds that the activation free energy
depends on distance through linear and quadratic terms which are
related through macroscopic arguments to surface and line tensions. We
find our present data set too sparse to elaborate on this finding.
At small separations, we include two states on the
wet surface for which the solvation state is dry for long times.  The
inclusion of such ``transient wet'' states, which may be visited as
the ligand diffuses along $q$, place the kinetics of assembly
predicted by the model in quantitative agreement with molecular
dynamics simulation.

The diffusion coefficient in the dewetting region is taken to be
constant, albeit with different values on the wet and dry surfaces
($D_\text{wet} = D_\text{dry}/2 = 7.56\times 10^{-4}\,
\text{nm}^2/\text{ps}$).  The value of $D_\text{wet}$ is taken to be
the diffusion constant for the ligand at large separations. In this
way, the hydrodynamic effect in the dewetting region is subsumed into
the wet-to-dry transitions which are an explicit degree of freedom of
the dynamics given by Eq. \ref{eq:motion_eq}.  Outside this
region, the diffusion coefficient is parameterized from the frictional
profile (Fig. \ref{fig:Layout_semi}A).

\begin{table} 
 \caption{The period of wet to dry
    transitions and equilibrium occupancy of wet and dry states. At small
    $q$,  no occupancy is given as it is dominated by the dry state.} 
\label{tab:wet2dry}
\begin{tabular}{ccccc} \hline $q$ & $k^{-1}(q, d \leftarrow w)$ &
$P_\text{dry}$ & $P_\text{wet}$ \\ \hline 0.036 nm & 6.0 ps & -- & --
\\ 0.086 nm & 20.0 ps & -- & -- \\ 0.136 nm & 75.8 ps& 0.74 & 0.26 \\
0.186 nm & 193 ps & 0.44 & 0.56 \\ 0.236 nm & 212 ps & 0.29 & 0.71 \\
0.286 nm & 276 ps & 0.38 & 0.62 \\ 0.336 nm & 519 ps & 0.17 & 0.83 \\
\hline
\end{tabular}
\end{table}

\subsection{Comparison of Diffusive Surface Hopping Model with other
  dynamical schemes} \label{sec:dynamics} 
In order to directly compare the dynamics of the DSHM to explicit
solvent molecular dynamics simulation and to one-dimensional
  Smoluchowski dynamics, we determine the time-dependent spatial
  distribution function, $P(q,t)$, and the mean first passage times
  (MFPT) for assembly from MD simulations where the pocket is fixed
  and the ligand is free to move in the direction of $q$. To guarantee
  that the ligand cannot diffuse far from the binding site, a
  repulsive wall potential is added to the system. The resultant
  potential of mean force, including the repulsive wall, is depicted
  in Fig. \ref{fig:mastereq}. It is important to note that these
  simulations differ from those from which the
 model parameters were determined, where $q$ was fixed at
 different values.

Apart from comparing with MD simulations, it is also of interest to
compare the two dimensional model (DSHM) to one-dimensional
diffusion (Smoluchowski) dynamics on the average potential of mean
force described by the Markovian master equation, Eq. (\ref{eq:smoul})
including either a position-dependent or a constant diffusion
coefficient~\cite{szabo98}.  The diffusive dynamics occurs on the
average potential of mean force that includes contributions from both
wet and dry states.  When hydrodynamic interactions are considered
(Avg-HI), the friction coefficient profiles depicted in
Fig.~\ref{fig:Layout_semi} are utilized.  In the case of no
hydrodynamic interaction (Avg-NOHI) the diffusion constant $D_\text{wet}$ 
is utilized for all values of $q$.

\begin{figure}[t]
\begin{center}
\includegraphics[scale=0.63]{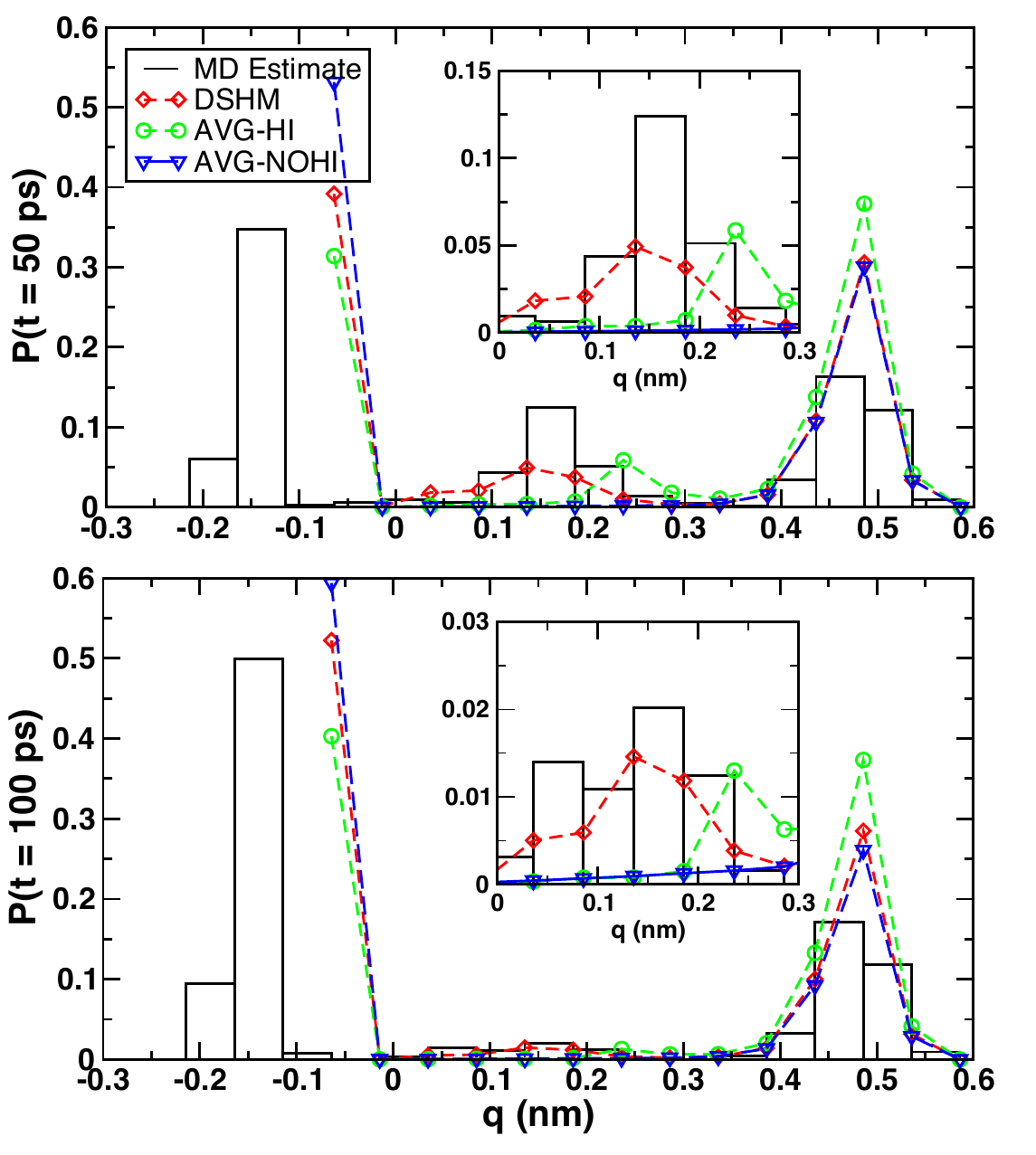}
\end{center}
\caption{Probability distribution of the assembly process at $t=50$ ps (top panel) and
$t=100$ ps (bottom panel). The results from MD simulation (black bars) are
compared against those extracted from three theories that can
be expressed as Markovian master equations, one where the distribution
evolves on two surfaces (DSHM, red line with diamonds), and where it evolves
on an average surface with (AVG-HI) and without hydrodynamic interactions (AVG-NOHI)
(green line with circles and blue line with triangles,
respectively).  Insets zoom into the distributions in the drying region.  } 
\label{fig:probpos}
\end{figure}

The spatial probability distribution at time $t$, $p(q,t) = \sum_s
p(s,q,t)$ can be compiled from a set of molecular dynamics
trajectories and compared to the results for the models that are
obtained from the solution of Eq. \ref{eq:mastereq}.  The probability distribution 
at $t=50$ ps and $t=100$ ps, is plotted
in Fig. \ref{fig:probpos}.  It can be readily seen that the MD
result exhibits three peaks: one corresponding to the basin in the
mean force that is created by the wall potential at large separations,
a smaller, more transient peak in the dewetting region, and a peak
corresponding to ligand in the docked state.  For long times, the
distribution is localized in the docked pose, or in the case of the
models based on Eq. (\ref{eq:mastereq}), in the absorbing state.

Whereas all Master Equation models considered reproduce the features
at the ligand far from the pocket and for the ligand in the docked
pose, the peak resulting from drying fluctuations is not described by
the average potential of mean force alone (which is strictly
  attractive in this region).  The results for MD and the DSHM model
are in good agreement in this region (see insets of Fig.
\ref{fig:probpos}), as the model captures the peak position and decay
from $t=50$ ps to $t= 100$ ps very well given the model's resolution.

The results for Avg-HI given in Fig.~\ref{fig:probpos} also exhibit
a peak in $p(t)$ in the drying region, but it is shifted with respect
to the results of the MD and the DSHM.  The peak in the Avg-HI distribution is
shifted to the right of where the friction coefficient peaks and is
related to where the smallest element of $R_{ij}$ appears in the
dewetting region (see the Appendix).  From consideration of
the DSHM, it is clear that this peak position is determined by the
local minimum along the wet surface and that it is successful in
reproducing the peak position present in the MD result.
This shift in peak position is reminiscent of
that reported in Ref. \onlinecite{setny13}, where a difference was observed in the $\zeta(q)$
computed from  simulations in which  $q$ is restrained,  and an effective spatial friction extracted from
MFPT data. The authors attributed this shift to non-Markovian effects which we have seen are captured by DSHM.

\begin{table*}[t] \caption{ Mean first passage times (MFPT) as
extracted from various models.} \label{tab:mpt}
\begin{tabular}{cccc} \hline & MFPT & MFPT & MFPT \\ &
$q_0=0.336$ nm & $q_0=0.286$ nm & $q_0=0.186$ nm \\ \hline 
MD & 473 ps & 45 ps & 42 ps \\ DSHM & 351 ps & 42 ps & 34 ps \\ 
Avg-HI& 816 ps & 52 ps & 5  ps \\ 
Avg-NOHI & 193 ps & 9 ps & 3 ps \\ \hline
\end{tabular}
\end{table*}

The mean first passage times (MFPT) for assembly initiated from a wet
state at $q_0=0.336$ nm are given in Table 2 for the models
considered.  It can be seen that, among all dynamical schemes
presently considered, the MFPT obtained from DSHM comes closest to the
results obtained from MD. The Avg-HI result significantly
overestimates the MFPT.  On the other hand the MFPT obtained from
Avg-NOHI is a drastic underestimation of the result obtained from
simulation, partly because it lacks a description of the drying
transition.

In order to gain a more detailed understanding of the kinetics in the
drying region, we compute the MFPT to assembly from initial wet
configurations at pocket-ligand separations where the average surface
points downhill towards assembly.  Trajectories (about $3\%$) that
recross into the region beyond the drying transition are not counted
for the purpose of this calculation.  The DSHM is in agreement with
the results obtained from MD.  As expected, the AVG-NOHI MFTP is far
too low owing to its lack of a description of the drying phenomena.
Interestingly, the AVG-HI model yields a reasonable result for one
initial condition ($q_0=0.286$ nm) but not the other ($q_0=0.186$ nm).
This result is another manifestation of the spatial shift of the
probability distribution discussed above, such that the large
hydrodynamic slowdown does not influence the (AVG-HI) results when the ligand
is initially placed to the left of the center peak in Fig.
\ref{fig:probpos}.

\section{Conclusions} \label{sec:conc} A full assessment of the
kinetics of molecular recognition processes calls for the inclusion of
molecular-scale hydrodynamic effects.  However, most typically in
coarse-grained models such effects are either ignored or treated
within Markovian limit where the solvent time scales are assumed
to be fast compared to those of the heavy bodies. In reality, however
slow solvent fluctuations are present when confined water molecules
are expelled from the region between the ligand and the pocket wall.
The non-Markovian nature of this problem begs for a more comprehensive
theory which includes the solvent as an explicit part of the reaction
path.

We present a simple model to study the kinetics of ligand binding in a
model hydrophobic enclosure in conjunction with a novel coarse-grained
theory for describing the solvent behavior in which solvent is
accounted for by introducing a discrete state variable specifying
whether the pocket is ``dry'' or ``wet.''  In this way, diffusive
motion along the (heavy body) assembly coordinate is coupled to
transitions between the wet and dry surfaces.  This model is found to
yield an improved description of the assembly process when compared
with models that ignore these state changes and obey standard
Smoluchowski (Brownian) dynamics.  In this way, the leading phenomena
that give rise to non-Markovian behavior may be subsumed into a
Markovian master equation description that lies in a larger state
space.

Here we have explored the role of solvent in the kinetics of
ligand-pocket association.  Although our model is rather crude, it is
still able to capture the displacement of water molecules by a ligand
via hydrophobic drying transitions and the free energy barrier
associated with ligand desolvation.  However the ligand pocket is
rather smooth and interactions are only mediated through the
Lennard-Jones potential and not specific hydrogen bonding.  Indeed,
the presence of water may be more or less favorable near hetergeneous
surfaces\cite{rossky07} or in different regions of the
pocket~\cite{young07}.  Furthermore, the pocket and the ligand are
both rigid structures in the simulations and the coupling of ligand
and pocket internal degrees of freedom is not presently considered.
Such effects may be rather slow and essential in the pathway to
assembly, and may be incorporated in modified diffusive surface
hopping models.

Assembly processes that occur in a bath of lighter particles are
important in a wide variety of settings.  Molecular dynamics
simulations provide an excellent tool for the study of such systems,
although a deep understanding can be gained from the relation of large
scale simulations to more coarse grained models of diffusion on
smaller subsets of collective variables.  This understanding can
further clarify the kinetics of assembly processes and be incorporated
into coarse-grained models.  Future work will address this relation,
as well as the behavior of patchy, softer, and more realistic bodies.

\begin{acknowledgments}
This work was supported by grants from the National
Institutes of Health [NIH-GM4330] and the National Science Foundation
through [via Grant No. NSF-CHE-0910943]. We gratefully acknowledge the
computational support of the Computational Center for Nanotechnology
Innovations (CCNI) at Rensselaer Polytechnic Institute (RPI). This work
used the Extreme Science and Engineering Discovery Environment
(XSEDE), which is supported by National Science Foundation grant
number OCI-1053575.
\end{acknowledgments}

\appendix
  \section{Expression for elements of dynamic matrix} 
\begin{figure}[h]
\begin{center}
\includegraphics[scale=0.25]{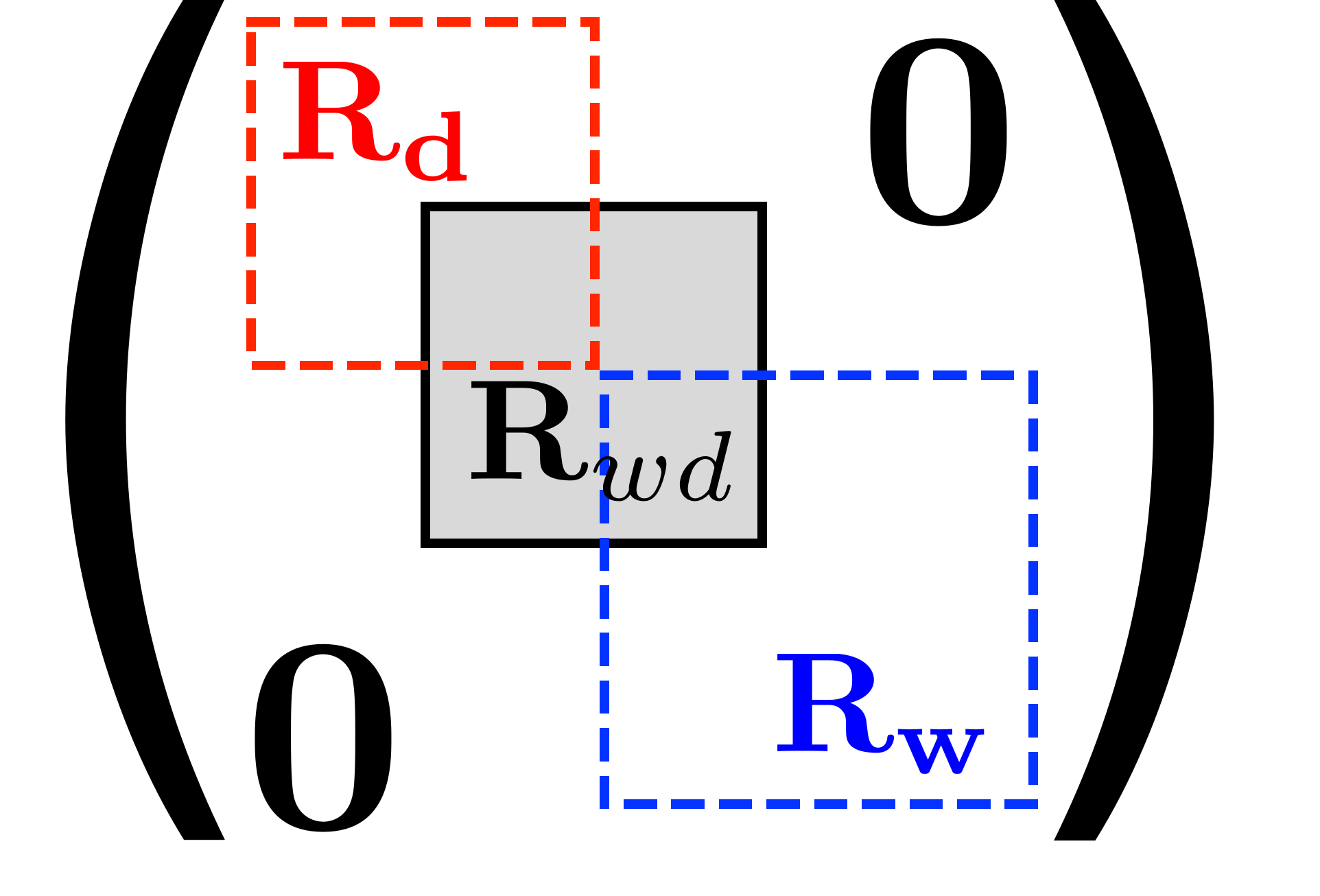}
\end{center}
\caption{A schematic of the addition of the matrices that comprise the
  rate matrix, $\mathbf{R}$.  Transitions along $q$ on the dry
  surface are described by $\mathbf{R}_d$, and along the wet surface
  by $\mathbf{R}_w$.  $\mathbf{R}_{wd}$ mediates transitions between
  the surfaces.  The non-zero blocks of these matrices are denoted by
  the red, blue, and black boxes, respectively.}
\label{fig:matrix}
\end{figure}

The dynamics of
  a system that obeys a Markovian master equation description is given
  by Eq. \ref{eq:mastereq} for the evolution of the probability
  distribution in state $i$, $p_i(t)$ In the two surface model, the
  index i runs over all allowed wet and dry states. The diffusive
  coordinate $q$ is represented on a grid of spacing $\Delta q$. The
  matrix $\mathbf{R}$ can be expressed as the sum of three matrices
  that describe distinct types of transitions,
\begin{equation} \mathbf{R} = \mathbf{R}_d + \mathbf{R}_w +
\mathbf{R}_{wd} \; . \label{eq:matsum}
\end{equation} 
A diagram of this addition is shown in Fig. \ref{fig:matrix}.  The total
dimension of $\mathbf{R}$ is $N_w + N_d$.  The matrices $\mathbf{R}_d$
and $\mathbf{R}_w$ describe transitions along $q$ in the dry and wet
states, respectively and have the following form:
\begin{widetext}
\begin{equation} R_s(q_i,q_j) = -\left[ \omega_{+,j}^{(s)} +
\omega_{-,j}^{(s)} \right] \; \delta_{ij} + \omega_{+,j}^{(s)} \;
\delta_{j+1,j} + \omega_{-,j}^{(s)} \; \delta_{j-1,j}
\end{equation}
where `$s$' denotes either a wet or dry state and,
\begin{eqnarray} \omega_{+,n}^{(s)} &=& \frac{D(s,q_{n+1})+D(s,q_n)}{2
\Delta q^2} \exp\left(+\alpha \right) \\ \omega_{-,n}^{(s)} &=&
\frac{D(s,q_{n-1})+D(s,q_n)}{2 \Delta q^2} \exp\left(-\alpha \right)
\; ,
\end{eqnarray} where,
\begin{equation} \alpha = \frac{\beta \Delta
q}{4}\left[\bar{F}(s,q_{n+1})+\bar{F}(s,q_n)\right] \; .
\end{equation}  $D$ and $\bar{F}$ are the diffusion
coefficient and mean force on a particular surface.  These expressions represent a
discretization of the Smoluchowski Equation (Eq. \ref{eq:smoul})~\cite{szabo98}.  
\end{widetext}
In the Markovian (Brownian) limit, this expression along a single,
averaged surface is considered.  The spatial dependence in the
diffusion coefficient engenders a maximum slowdown in the transition
probability $R_{ij}$ where the sum of the diffusion coefficients in
states $i$ and $j$ is minimum. For the profile plotted in Fig.
\ref{fig:Layout_semi} this occurs at $q= 0.236$ nm, and is reflected
in the peak position of $p(q,t)$ that is observed in the AVG-HI model
(Fig. \ref{fig:probpos}).  The peak position also can depend on
$\bar{F}$, but it is roughly constant in the dewetting region on the
average surface.

In the range of separations where wet-to-dry transitions may occur,
they are described by the matrix $\mathbf{R}_{wd}$.  If $N_{wd}$ is
the number of values of $q$ for which such transitions are allowed,
then the matrix is (square) block diagonal where the block has
dimension of $2N_{wd}$.  This block has diagonal elements: $-k(q,w
\leftarrow d)$ for indices less than or equal to $N_{wd}$ and $-k(q,d
\leftarrow w)$ for indices greater than $N_{wd}$.  The off diagonal
components are non-zero for transitions between wet and dry states at
the same value of $q$.  The lifetime of the wet state is the inverse
of the transition rates given in Table 1.

\section{Simulation model and details} 
\label{sec:methods}
\begin{figure}[t]
\includegraphics[width=3.0in]{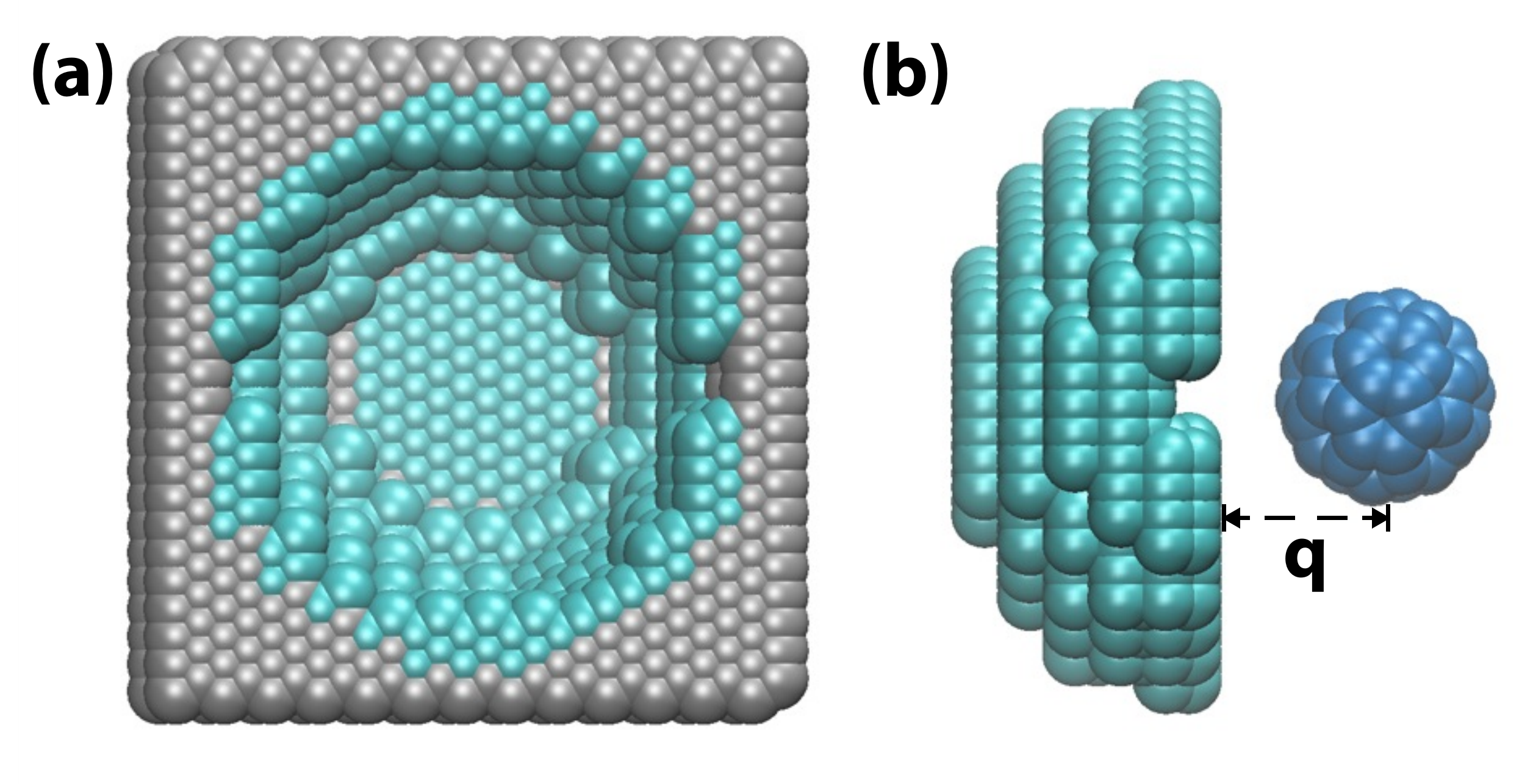}
\caption{Front and side view of the model ligand (C60 molecule, blue)
and model pocket are depicted in panels (a) and (b), respectively.
The well-depth associated with the LJ interaction ($\epsilon$) of
sites that line the pocket (green atoms) is varied in this study.
Other sites (gray atoms) are assigned the smallest value of $\epsilon$
utilized in this work.  The model is visualized with VMD.}
\label{fig:model}
\end{figure}

The model system in the current study, as depicted~\cite{vmd96} in
Fig. \ref{fig:model}, is inspired by the one developed by McCammon
and coworkers,~\cite{Baron:JACS10} but with some distinct features, 
most notably that the size of the ligand and
the pocket are larger, with a length scale on the order of 1 nm.
In the present model, a semi-ellipsoidal (rather than hemispherical)
pocket is ``cut out'' of a hexagonal close-packed (HCP) lattice. The
equation for the ellipsoid is:
\begin{equation} x^2 + y^2 + (cz)^2 = R^2 \; ,
\end{equation} with c=0.8528 and R=1.1 nm.  The pocket sites are fixed
and interact with the model ligand with an Lennard-Jones potential
with $\sigma$= 0.4152 nm, as in Ref ~\cite{Baron:JACS10}.  The well
depth is varied for the sites that line the pocket, as determined by a
thickness of 0.17 nm.  In this way, we vary the extent of
hydrophobicity of the pocket. We have utilized three pocket types with
different interaction potential well-depths: a) ``highly attractive''
interaction with $\epsilon$=0.024 kJ/mol, b) ``weakly attractive''
interaction with $\epsilon$=0.0024 kJ/mol and c) ``intermediate
attractive'' interaction with $\epsilon$=0.008 kJ/mol . All other
pocket sites are assigned an $\epsilon=$ 0.0024 kJ/mol as in
Ref. \onlinecite{Baron:JACS10}.  The main text concentrated on the case of
intermediate attractive attraction (pocket $\epsilon$=0.008 kJ/mol )
and here, we elaborate on the other two parameter choices, namely the
``highly attractive'' and ``weakly attractive'' pocket.

Unlike in the work of McCammon and
co-workers~\cite{Baron:JACS10,setny13} where a methane molecule was
used to represent the ligand, we employ a nanoscale object in the form
of C60 fullerene.  This model of C60 fullerene has been utilized in
one of our prior studies~\cite{joe12} and interacts with other bodies
by means of a Lennard Jones potential with parameters $\epsilon$=0.276
kJ/mol and $\sigma$=0.350 nm. The water is modeled with the TIP4P
potential~\cite{tip4p}. The solute-solvent interactions are represented
by the geometric mean of the respective water and solute parameters.
 
 \section{Technique to compute the hydrodynamic profile}
  
In the Brownian limit, one can quantify the solvent-induced potential
of mean force (PMF) and hydrodynamic profile between the pocket and
ligand along a coordinate that is a function only of heavy-body
positions.  In the present work, this coordinate is along the
direction of the difference between the center-of-masses of the pocket
and the ligand and is zero where the ligand center of mass enters the
cavity (see Fig. \ref{fig:model}B)).  This reaction coordinate,
$q$, is only a function of the heavy bodies.  The effect of the water
is included implicitly through the induced potential of mean force
experienced by the bodies and the hydrodynamic interactions encoded in
the friction coefficient.  The non-Markovian effects that arise from
the slow solvent fluctuations are not treated by this framework, and
the development of an alternative model (i.e. DSHM) is one of the
chief aims of this study.

To calculate the solvent-induced potential of mean force, we utilize a
similar protocol as in our prior work~\cite{joe12,jingyuan12}.  A
series of simulations are run keeping the pocket and ligand fixed at
selected values of $q$.  The ligand-pocket separation ranged from 0.86
nm to 2.46 nm at a separation of 0.1 nm and where needed, finer grid
of spacing 0.05 nm was also used.  The simulation box dimension, prior
to equilibration, varied between 4.6 nm $\times$ 4.6 nm $\times$ 6.0
nm and 4.6 nm $\times$ 4.6 nm $\times$ 8.0 nm, depending on the
pocket-ligand separation. Accordingly, the number of solvent molecules
in the box also varied between 3375 and 4774. At each separation, the
system was equilibrated under isothermal-isobaric (NPT) conditions for
1 ns and the data were collected in the canonical (NVT) ensemble for 5
ns.  A stochastic velocity rescaling thermostat~\cite{Bussi:2007p114}
was employed to maintain the temperature at 300 K for all simulations
a Berendsen barostat~\cite{berendsen} was employed at a pressure of 1
bar for constant pressure equilibration.  The solvent-induced mean
force acting on the ligand and pocket are then computed from NVT runs
at each separation. To obtain the solvent-induced potential of mean
force (PMF), this quantity is then integrated along the reaction
coordinate. Simulations were performed in the microcanconical (NVE)
ensembles in order to evaluate the pocket-water fluctuations and
solvent relaxation time. To maintain strict energy conservation in
the microcanonical ensemble, NVE runs were performed in double
precision.  All simulations were performed using GROMACS
4.5.4~\cite{gromacs4}.

The position-dependent relative friction coefficient acting on the
ligand as a function of pocket-ligand separation is computed in the
following manner. In our recent work the friction coefficient was
computed from the force-force autocorrelation
function,~\cite{Bocquet:1997p103} but in the present study we utilize
an alternative method \cite{Straub:1990p65,hummer05} to compute the
friction coefficient.  In this method, one performs series of umbrella
sampling simulations along the reaction coordinate and extracts the
friction from the time correlation function of reaction coordinate
position at a given separation:
 \begin{equation} \zeta(q) = k_bT \; \frac{\int \mathrm{d} q
\left<\delta q(t) \delta q(0) \right>}{\left<\delta
q^2\right>^2} \; , \label{eq:friction}
\end{equation} where $\delta q = q-\left<q\right>$ represents the
fluctuations of the reaction coordinate.

In our current study, the umbrella sampling technique is utilized
instead of the force-force autocorrelation technique
\cite{Bocquet:1997p103} that we previously
employed~\cite{joe12,jingyuan12}. This is mainly due to the fact that
the symmetries of the two-body friction tensor that are exploited in
the work of Ref. \onlinecite{Bocquet:1997p103} are not present when the two
bodies are not identical as in the pocket-ligand complex.

The protocol for the friction calculation is as follows.  The force
constant of the umbrella potential was chosen such that it was the
lowest value that engenders a Gaussian distribution about the average
value of the reaction coordinate. The typical values of the force
constant for the umbrella potential ranged between $3000 - 6000$
kJ/mol/nm$^2$.  At the ligand-pocket separation corresponding to the
friction peak, we also repeated the calculation of the friction
coefficient using different force constants.  The uncertainties in the
friction coefficient computed using different force constants was less
than 10\% of the average values. The following protocol was employed
in order to compute position time correlation function to good
statistical accuracy: At selected ligand-pocket separations, the
system is first subjected to a short umbrella-sampling equilibration
in the canonical (NVT) ensemble. Subsequently, microcanonical (NVE)
umbrella sampling production runs were initiated by varying the
initial velocity seeds.  In the dewetting region, where very slow
solvent fluctuations are present, twenty independent runs of up to 8
ns were carried out. Outside this region, it was found that ten, 4 ns
long trajectories are sufficient to converge the results. The values
of instantaneous reaction coordinates were collected every 0.01 ps
from each run.  To facilitate energy conservation, double precision
routines and a smaller time step of 0.001 ps was utilized. Due to the
expensive nature of the computation, the friction coefficient was
computed at selected values of reaction coordinate. An estimate of the
error in friction coefficient was also obtained by block averaging
over multiple umbrella sampled trajectories for largest ligand-pocket
separations as well as the ligand-pocket separation corresponding to
friction peak. The uncertainty in the friction coefficient was 1-3$\%$
of the average values for the highly attractive and intermediate
attractive cases and 1-15$\%$ of the average values for the weakly
attractive cases.

For the purpose of analysis, the variation of pocket-water and
ligand-water as a function of distance are also computed.
The `ligand-water' is defined as the water in the first solvation
shell ($R_\text{CF-OW}< 0.8$ nm) of the ligand, and the `pocket water'
is simply the number of water molecules present in the pocket pocket
(see Fig \ref{fig:model}). The fluctuations and relaxation
time of the pocket waters are computed at different ligand-pocket
separations. The solvent fluctuations are given by the expression
$\left<\delta N^2\right>/ \left< N \right>$ and the relaxation time
by,
\begin{equation} \tau = \frac{1}{\left<\delta N^2 \right>}\int
\mathrm{d}t \left< \delta N(t) \delta N(0) \right> \; ,
\end{equation} where $N$ is the number of water molecules in the
pocket and $\delta N = N - \left<N\right>$.  In our previous work,
such properties have been found to yield trends relatable to those
observed for the profile of the friction
coefficient~\cite{joe12,jingyuan12}.

\section{Hydrodynamics and solvent fluctuations in the hydrophilic and
ideally hydrophobic cavities}

\begin{figure}[t]
\begin{center}
\includegraphics[width=3.0in]{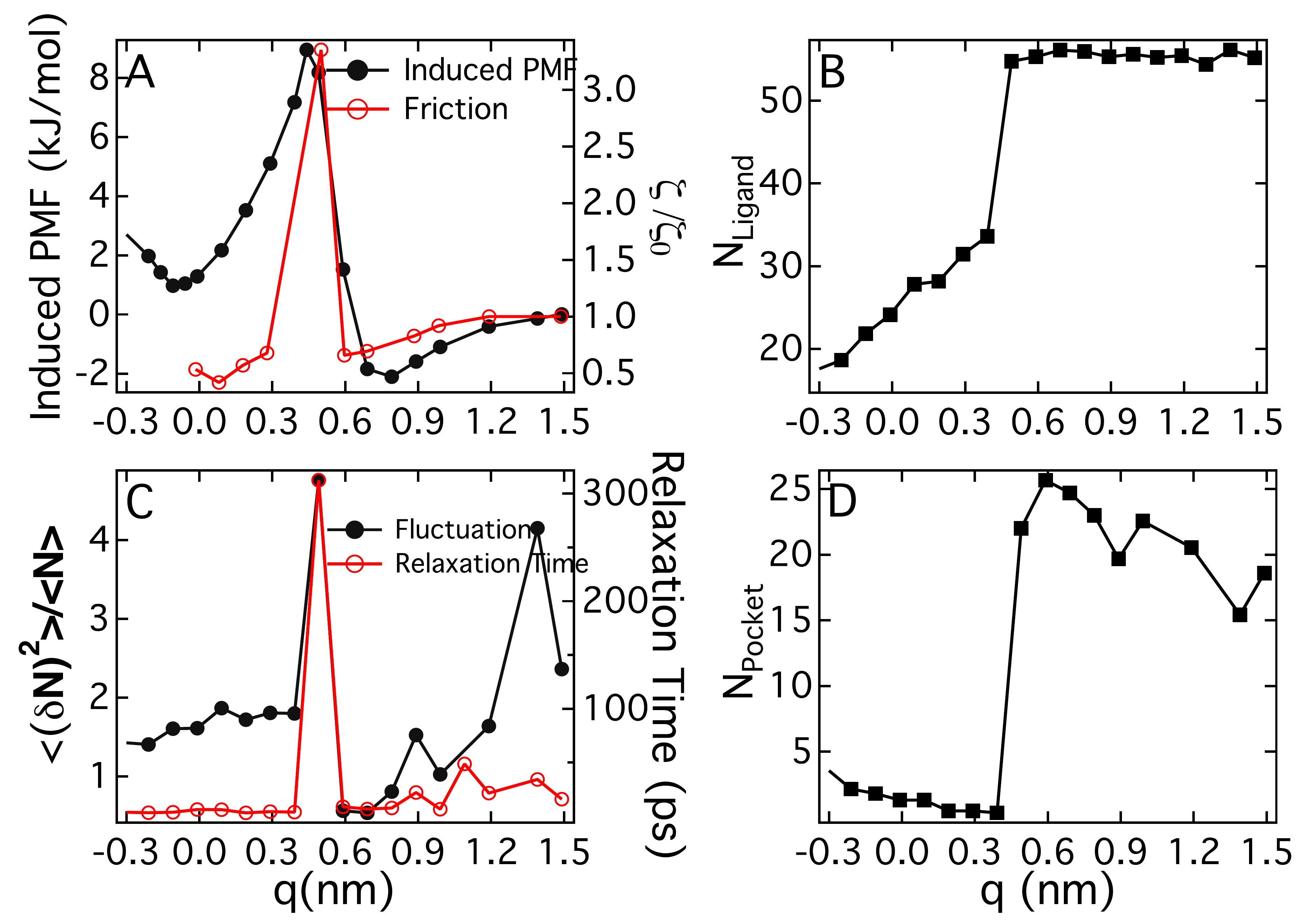}
\end{center}
\caption{Layout describing the correlations among various
  thermodynamic and hydrodynamic profiles as a function of
  ligand-pocket distance for weakly attractive pocket. (A) Comparison of
  the solvent-induced PMF (left scale) and frictional profile (right scale), (B) 
  ligand-water number, (C) correlation of
  pocket-water fluctuation (left scale) and relaxation time (right scale), and (D) pocket-water number.}
\label{fig:layout_weakly}
\end{figure}

\begin{figure}[t]
\begin{center}
\includegraphics[width=3.0in]{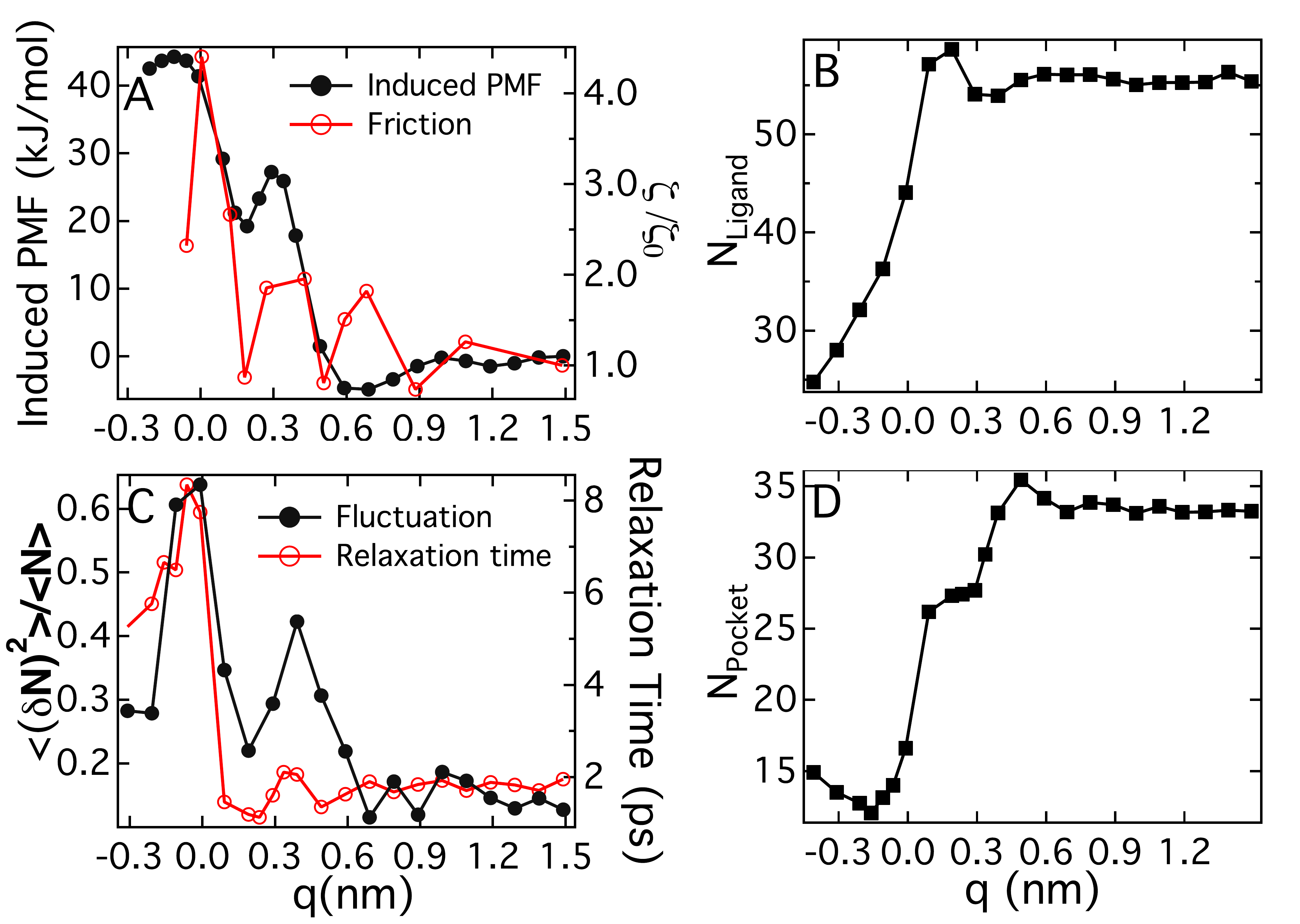}
\end{center}
\caption{Layout describing the correlations among various
  thermodynamic and hydrodynamic profiles as a function of
  ligand-pocket distance for highly attractive pocket. (A) Comparison of
  the solvent-induced PMF (left scale) and frictional profile (right scale), (B) 
  ligand-water number profile, (C) correlation of
  pocket-water fluctuation (left scale) and relaxation time (right scale), and (D) pocket-water number profile.}
\label{fig:layout_attract}
\end{figure}

The hydrodynamic profiles and the corresponding water-profile for the
intermediate attractive pocket ($\epsilon$=0.008 KJ/mol) have been
discussed in the main text. In this Appendix, we will
primarily discuss the results for the two remaining cases, namely the
highly and weakly attractive pocket. In Figs.
\ref{fig:layout_weakly} and \ref{fig:layout_attract}, we present
pocket-water, ligand-water, the potential of mean force and variation
of the frictional coefficient with separation for the ligand
interacting with the weakly and highly attractive pockets,
respectively.

It can be seen that, even when the ligand is far away from the pocket
(for large values of $q$), different water densities are present in
all three pockets presently considered (see the main text for the
pocket of intermediate attraction), owing to differences in the pocket
hydration affinity.  Naturally, as the C60 potential is not modified,
the number of waters in the ligand solvation shell is the same at
large $q$.  As the ligand approaches the pocket, water is stripped
from its solvation shell, and as it is rather hydrophilic, one
can expect a contribution to the barrier to assembly arising from this
process.  There is an observed maximum in the number of pocket water
as the ligand approaches the cavity, owing to the overlap of ligand
solvation shell into the pocket.

When the ligand starts to enter the pocket ($\approx q= 0.5$ nm), the
weakly attractive pocket undergoes a drying transition (see
Fig. \ref{fig:layout_weakly}B).  As in the case of intermediate
attraction discussed in the main text, high friction coefficients and
large and slow solvent fluctuations are associated with the drying
transition, albeit at a larger pocket-ligand separation.  The
solvent-induced free energy profile includes a relatively shallow
basin at $q=0.8$ nm and a barrier that is primarily engendered by the
desolvation of the ligand.  Hydrophobic attractive forces dominate and
drive the ligand pocket complex to assembly, but there is an
unexpected turnover of the free energy profile at very small pocket-ligand
separations near the binding pose. We attribute this increase in the
PMF for $q<0$ to the retention of water brought in by the hydrophilic
ligand, which is unfavorable in this hydrophobic pocket (see Fig
\ref{fig:layout_weakly}D).  We also note that even though most of
these features in the solvent-induced free energy profiles are also
observed in the case of the pocket with intermediate attractions, the
above-mentioned increase in free energy near the binding position is
absent.

In the highly attractive pocket (Fig. \ref{fig:layout_attract}),
solvation water is stabilized by the relatively strong affinity of
water for the cavity, even as the model ligand penetrates deeply into
the pocket. Peaks in the friction correspond to features in the
potential of mean force for $q<0.5$ nm, as well as to large solvent
fluctuations and relaxation times.  The molecular-scale hydrodynamic
effects are a manifestation of slow, confined water molecules which
are ``squeezed out'' as the ligand is brought closer to the cavity
wall. The highly attractive pocket expels water layer-by-layer as
indicated by the decreases in number at $q=0.3$ nm and $q=0$.  The
difference in these separations is dictated by the diameter of the
water molecule ($\sigma_\text{O}\approx 0.3$ nm). As shown in our
previous work~\cite{joe12,jingyuan12}, the peaks in the friction can
fortify the kinetic barriers to association that the ligand
experiences as it traverses the free energy surface.

\section{Further description of the Diffusive Surface Hopping Model}

We propose a two dimensional model where one coordinate is the
diffusive coordinate $q$ (the separation between receptor and ligand),
and the other is a discrete state variable $s$, indicating whether the
binding pocket is wet or dry. As shown in the Eq. 2 in the main text,
this model has state dependent diffusion constants $D(s,q)$, evolves
on state dependent free energy surfaces $W(s,q)$, and its state can
change from wet to dry and from dry to wet by first order kinetics
with rate constants also depending on $q$.  This model is equivalent
to a particle diffusing on a potential energy surface that can hop
between two functional forms, one for the wet and one for dry states,
with different spatially dependent diffusion constants on each
surface. We call this the Diffusive Surface Hopping Model (DSHM). The
diffusion dynamics is thus coupled to transitions between the surfaces
specified with rate constants $k(s\leftarrow s^\prime,q)$.

In our present model, 22 (9 dry and 13 wet) states are employed with a
grid spacing of $\Delta q = 0.05$ nm to describe the system on the
domain of $-0.064 \text{ nm} \le q \le 0.636 \text{ nm}$.  For the
purpose of this model, a state is defined as dry if there are fewer
than 15 water molecules in the pocket, and wet otherwise.  The mean
force is extracted from simulations where the ligand and pocket are
fixed, and averages are taken separately over wet and dry states. The
potentials of mean force corresponding to the wet and dry states are
plotted in Fig. 3 in the main text.  These surfaces include the
direct interactions of the heavy bodies, which are strictly attractive
on the domain of $q$ that is currently considered.  It can be seen
that there is a relatively shallow minimum at $q\approx 0.15$ nm. The
dry surface is strictly attractive and assembly proceeds readily after
a transition to a dry state occurs.  Transitions between the wet and
dry surfaces are allowed in the region $0.036 \text{ nm} \leq q \leq
0.336 \text{ nm}$.  The values of the transition rate constant at
allowed values of $q$ are given in Table 2 in the main text. Matrix
elements for transitions, $\mbox{wet}\rightleftharpoons\mbox{dry}$,
are estimated from the average dwell time in the dry state from MD
simulations at fixed $q$.  The transition times become shorter as the
bodies approach each other and fewer water molecules are displaced by
the drying event.  In general, the dry state becomes more favorable at
smaller separations. At $q=0.036$ nm and $q=0.086$ nm, the transition
rate parameters are estimated from the relaxation time of the solvent
to irreversibly go from the wet to dry state, as wet states are not
observed at long times.
  
In the drying region, the diffusion coefficient is approximated to be constant on each surface, 
albeit it is taken to be twice as large on the wet surface.
This ratio is estimated from the position-position autocorrelation
functions depicted in Fig. 3C of the main text. The value of the
diffusion coefficient on the wet surface was taken to be equal to the
diffusion coefficient at largest separations, $D_\text{wet} = 7.56\times
10^{-4} \, \text{nm}^2/\text{ps}$.  As the DSHM only explicitly
describes the wet to dry transitions, other hydrodynamic effects, such
as the confined nature of pocket water are not included, and therefore
this value may be considered an upper bound on an appropriate estimate
for $D_\text{wet}$.  Outside the drying region, the diffusion coefficient is parameterized from the frictional profile (see Fig. 2A 
in the main text).

An absorbing boundary is set at $q=-0.064 $ nm and a reflecting
boundary is present at the wet state value of $q=0.036 $ \text{nm}.
The former denotes a position where assembly is considered to occur
and the latter choice is justified by the fact no wet states are
allowed at smaller separations due to steric hinderance as observed
from the highly attractive case.

\section{Calculation of MFPT from MD simulations} In order to directly
compare the dynamics of the Diffusive Surface Hopping Model to those
of the explicit solvent molecular dynamics simulation and one
dimensional Smoluchowski Dynamics, we study the time-dependent spatial
distribution and the mean first passage times (MFPT) for assembly.
For this purpose, we have carried out a separate set of MD
  simulations, where the pocket is fixed and the ligand is free to
  move in the direction of $q$. To guarantee that the ligand cannot
diffuse far away from the binding site, a repulsive wall potential is
added to the system.  This is achieved by means of the PLUMED plugin
for GROMACS~\cite{plumed,gromacs4} and ensures that all configurations
are bound to the domain $q < 0.6$ nm. The functional form of the
boundary potential is as follows:
\begin{eqnarray} V_\text{wall}(q) & = & K(q-L)^{4} \;\;\;\;\; q \geq L
\nonumber \\ & = & 0 \text{ otherwise,}
\end{eqnarray} where $K = 15000$ kJ/mol/nm$^4$ and L is the lower
bound on $q$ which is set at $q=0.366$ nm (a position slightly to the
right of barrier in the total PMF).  With this boundary in place, the kinetics of assembly upon 
ligand entry into the pocket is
   described, but not the diffusion of the ligand
  up to the pocket entry.  A total of 645 different ligand-binding
trajectories were run, either by varying the velocity seeds or by
changing the solvent configurations at a fixed initial condition,
$q_0$. The initial configuration for each run is always chosen such
that the occupation of pocket waters corresponds to a wet state.  If a
dry state is chosen then assembly proceeds rapidly.

\begin{figure}[t]
\begin{center}
\includegraphics[width=2.8in]{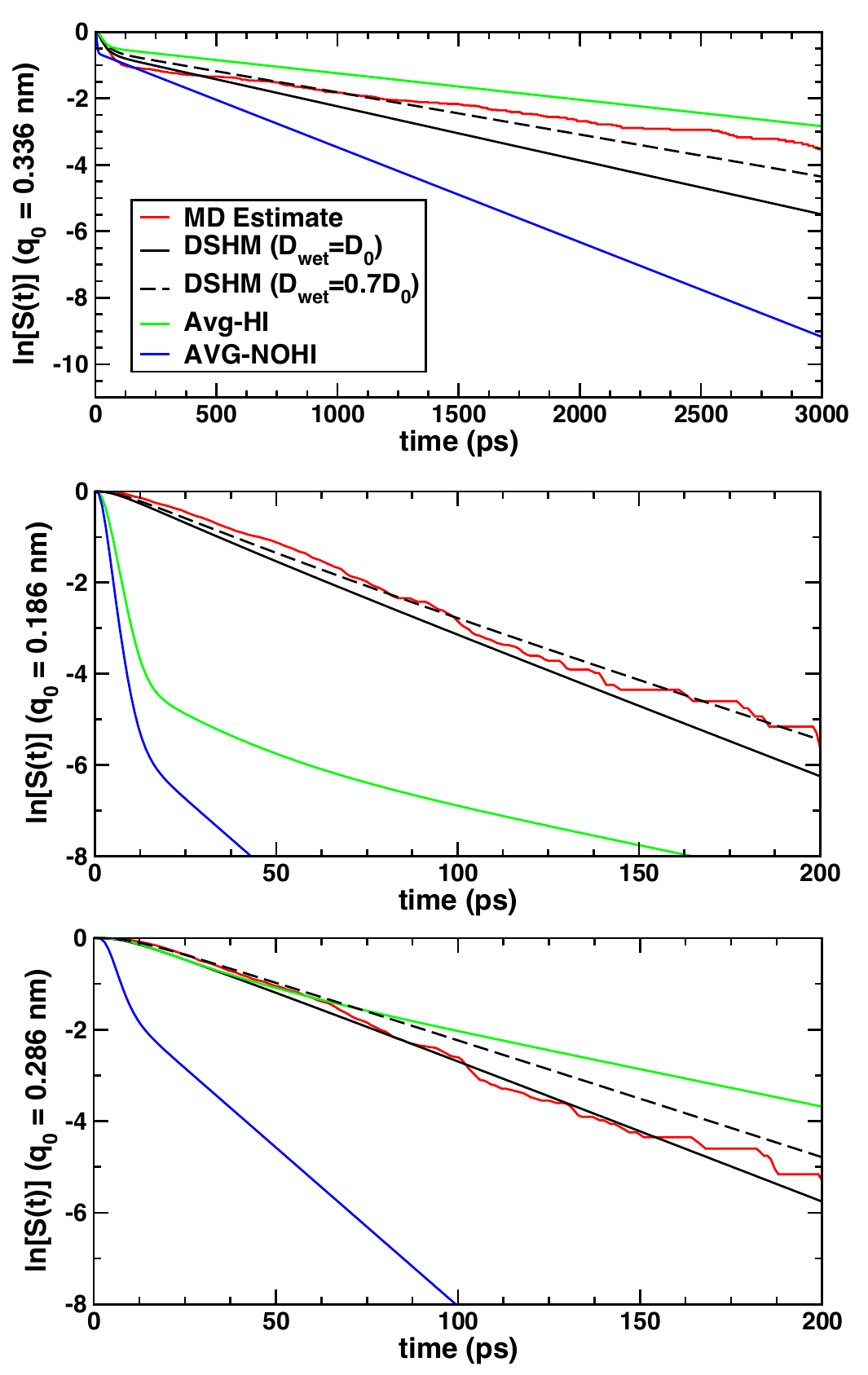}
\end{center}
\caption{The natural logarithm of the survival probability for the
assembly process as estimated from a set of explicit solvent molecular
dynamics trajectories (red curve) and coarse-grained models: the DSHM
for two choices of the wet state diffusion coefficient (black curves),
and Smoluchowski dynamics on the average surface with and without HI
(green and blue curves, respectively).  The data depicted in the top,
center, and bottom panels correspond to the MFPTs presented in Table 2
of the main text at the corresponding initial states.}
 \label{fig:survprob}
\end{figure}

\section{Survival probabilities: comparison of theory and MD}
The Markovian master equation is given by the following
expression,
\begin{equation} \frac{\partial p_i(t)}{\partial t} = R_{ij} p_j(t) \;
,
\end{equation} and has the following formal solution,
\begin{equation} p_i(t) = T_{ij}(t) p_j(0) \; ,
\end{equation} where the matrix $\mathbf{T}$ is the propagator $e^{t
\; \mathbf{R}}$.  This exponentiated matrix may be readily evaluated
in the eigenbasis of $\mathbf{R}$, which may be computed by means of
standard linear algebra libraries.  Given an absorbing state (at
$i=1$), the survival probability may be computed from the sum over the
probability of residing in non-absorbing states
$S(t)=\sum_{i=2}^{N_w+N_d} p_i(t)$.  The mean passage time is then
obtained by taking the time integral of $S(t)$.

The survival probability of assembly from $q_0=0.336 \text{ nm}$ is
plotted in Fig. \ref{fig:survprob} on a log scale for the MD
estimate as well as for the DSHM and Smoluchowski dynamics with and
without hydrodynamic interactions (HI) on the average surface.  It can
be seen that although the DSHM exhibits reasonable agreement for the
MFPT (see Table 2 of the main text), there are discrepancies at both
short and long times.  At short times, this is due to the fact that a
larger fraction of MD trajectories initially proceed in the direction
of assembly ($67\%$) than predicted by the DSHM ($54\%$).

At longer times, the disagreement can be alleviated by using a
different choice for the diffusion coefficient on the wet surface in
the drying region, $D_\text{wet}$.  As discussed above, the value
chosen is that of the diffusion coefficient at large separations
(denoted as $D_0$ in Fig. \ref{fig:survprob}) and can be considered
an upper bound for this parameter.  If $D_\text{wet}$ is tuned so as
to yield the same MFPT as the MD estimate, then $D_\text{wet}$ is
found to be approximately $0.7D_0$. This result is also plotted in
Fig. \ref{fig:survprob}, and this alteration is shown to improve
agreement with the MD estimate at long times.

\begin{figure}
\begin{center}
\includegraphics[width=2.8in]{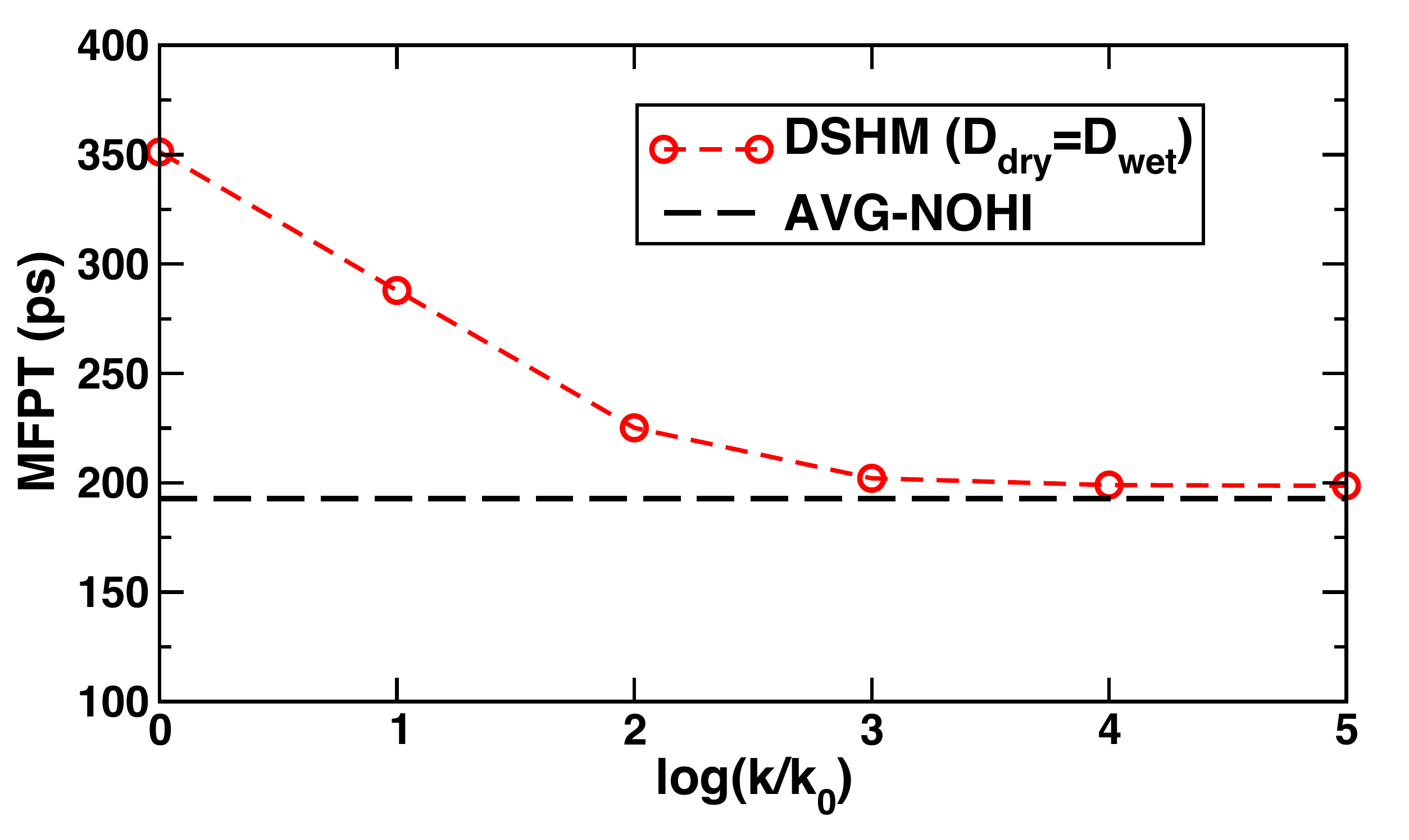}
\end{center}
\caption{The mean first passage time, is plotted for the DSHM as the
dewetting transition rates, $k$, are increased for their baseline
value, $k_0$ (red circles with dashed line).  For very fast
transitions, the results converge to the AVG-NOHI result (dashed
black line), when the diffusion coefficient is taken to be
$D_\text{wet}$ on both surfaces.}
\label{fig:survfast}
\end{figure}

In the middle and lower panels of Fig. \ref{fig:survprob}, we plot
the survival probably on a log scale that correspond to the passage
through the dewetting region and whose integral corresponds to the
MFPT values given the two right-hand columns of Table 2 in the main
text.  One can see how the Avg-HI model is far more sensitive to the
change in initial position than the other schemes.  The DSHM results
exhibit some sensitivity to the choice of $D_\text{wet}$, although
overall agreement with the MD estimate is not greatly altered.

As the transition rate between wet and dry states increases, the
resultant behavior should converge onto the result obtained from the
average surface.  This behavior arises from the fact that the
Smoluchowski dynamics is recovered in the limit where the solvent
dewetting fluctuations are fast.  Our model is indeed able to produce
this effect as is shown in Fig. \ref{fig:survfast}.  For the purpose
of this result, we only consider a single diffusion constant on both
the wet, dry, and average surfaces, $D_\text{dry} = D_\text{wet} =
D_\text{avg}=7.56 \times 10^{-4} \text{ nm}^2/\text{ps}$.  The initial
distribution is again taken to be a wet state at $q_0=0.336$ nm.  It
can be clearly seen that, given the chosen parameters, as the
transition rates are increased, the AVG-NOHI result is recovered from
the DSHM model.


\begin{thebibliography}{33}%
\makeatletter
\providecommand \@ifxundefined [1]{%
 \@ifx{#1\undefined}
}%
\providecommand \@ifnum [1]{%
 \ifnum #1\expandafter \@firstoftwo
 \else \expandafter \@secondoftwo
 \fi
}%
\providecommand \@ifx [1]{%
 \ifx #1\expandafter \@firstoftwo
 \else \expandafter \@secondoftwo
 \fi
}%
\providecommand \natexlab [1]{#1}%
\providecommand \enquote  [1]{``#1''}%
\providecommand \bibnamefont  [1]{#1}%
\providecommand \bibfnamefont [1]{#1}%
\providecommand \citenamefont [1]{#1}%
\providecommand \href@noop [0]{\@secondoftwo}%
\providecommand \href [0]{\begingroup \@sanitize@url \@href}%
\providecommand \@href[1]{\@@startlink{#1}\@@href}%
\providecommand \@@href[1]{\endgroup#1\@@endlink}%
\providecommand \@sanitize@url [0]{\catcode `\\12\catcode `\$12\catcode
  `\&12\catcode `\#12\catcode `\^12\catcode `\_12\catcode `\%12\relax}%
\providecommand \@@startlink[1]{}%
\providecommand \@@endlink[0]{}%
\providecommand \url  [0]{\begingroup\@sanitize@url \@url }%
\providecommand \@url [1]{\endgroup\@href {#1}{\urlprefix }}%
\providecommand \urlprefix  [0]{URL }%
\providecommand \Eprint [0]{\href }%
\providecommand \doibase [0]{http://dx.doi.org/}%
\providecommand \selectlanguage [0]{\@gobble}%
\providecommand \bibinfo  [0]{\@secondoftwo}%
\providecommand \bibfield  [0]{\@secondoftwo}%
\providecommand \translation [1]{[#1]}%
\providecommand \BibitemOpen [0]{}%
\providecommand \bibitemStop [0]{}%
\providecommand \bibitemNoStop [0]{.\EOS\space}%
\providecommand \EOS [0]{\spacefactor3000\relax}%
\providecommand \BibitemShut  [1]{\csname bibitem#1\endcsname}%
\let\auto@bib@innerbib\@empty
\bibitem [{\citenamefont {Young}\ \emph {et~al.}(2007)\citenamefont {Young},
  \citenamefont {Abel}, \citenamefont {Kim}, \citenamefont {Berne},\ and\
  \citenamefont {Friesner}}]{young07}%
  \BibitemOpen
  \bibfield  {author} {\bibinfo {author} {\bibfnamefont {T.}~\bibnamefont
  {Young}}, \bibinfo {author} {\bibfnamefont {R.}~\bibnamefont {Abel}},
  \bibinfo {author} {\bibfnamefont {B.}~\bibnamefont {Kim}}, \bibinfo {author}
  {\bibfnamefont {B.~J.}\ \bibnamefont {Berne}}, \ and\ \bibinfo {author}
  {\bibfnamefont {R.~A.}\ \bibnamefont {Friesner}},\ }\href@noop {} {\bibfield
  {journal} {\bibinfo  {journal} {Proc. Natl. Acad. Sci. USA}\ }\textbf
  {\bibinfo {volume} {104}},\ \bibinfo {pages} {808} (\bibinfo {year}
  {2007})}\BibitemShut {NoStop}%
\bibitem [{\citenamefont {Shan}\ \emph {et~al.}(2011)\citenamefont {Shan},
  \citenamefont {Kim}, \citenamefont {Eastwood}, \citenamefont {Dror},
  \citenamefont {Seeliger},\ and\ \citenamefont {Shaw}}]{shan11}%
  \BibitemOpen
  \bibfield  {author} {\bibinfo {author} {\bibfnamefont {Y.}~\bibnamefont
  {Shan}}, \bibinfo {author} {\bibfnamefont {E.~T.}\ \bibnamefont {Kim}},
  \bibinfo {author} {\bibfnamefont {M.~P.}\ \bibnamefont {Eastwood}}, \bibinfo
  {author} {\bibfnamefont {R.~O.}\ \bibnamefont {Dror}}, \bibinfo {author}
  {\bibfnamefont {M.~A.}\ \bibnamefont {Seeliger}}, \ and\ \bibinfo {author}
  {\bibfnamefont {D.~E.}\ \bibnamefont {Shaw}},\ }\href@noop {} {\bibfield
  {journal} {\bibinfo  {journal} {J. Am. Chem. Soc.}\ }\textbf {\bibinfo
  {volume} {133}},\ \bibinfo {pages} {9181} (\bibinfo {year}
  {2011})}\BibitemShut {NoStop}%
\bibitem [{\citenamefont {Dror}\ \emph {et~al.}(2011)\citenamefont {Dror},
  \citenamefont {Pan}, \citenamefont {Arlow}, \citenamefont {Borhani},
  \citenamefont {Maragakis}, \citenamefont {Shan}, \citenamefont {Xu},\ and\
  \citenamefont {Shaw}}]{dror2011}%
  \BibitemOpen
  \bibfield  {author} {\bibinfo {author} {\bibfnamefont {R.~O.}\ \bibnamefont
  {Dror}}, \bibinfo {author} {\bibfnamefont {A.~C.}\ \bibnamefont {Pan}},
  \bibinfo {author} {\bibfnamefont {D.~H.}\ \bibnamefont {Arlow}}, \bibinfo
  {author} {\bibfnamefont {D.~W.}\ \bibnamefont {Borhani}}, \bibinfo {author}
  {\bibfnamefont {P.}~\bibnamefont {Maragakis}}, \bibinfo {author}
  {\bibfnamefont {Y.}~\bibnamefont {Shan}}, \bibinfo {author} {\bibfnamefont
  {H.}~\bibnamefont {Xu}}, \ and\ \bibinfo {author} {\bibfnamefont {D.~E.}\
  \bibnamefont {Shaw}},\ }\href@noop {} {\bibfield  {journal} {\bibinfo
  {journal} {Proc. Natl. Acad. Sci. USA}\ }\textbf {\bibinfo {volume} {108}},\
  \bibinfo {pages} {13118} (\bibinfo {year} {2011})}\BibitemShut {NoStop}%
\bibitem [{\citenamefont {Copeland}\ \emph {et~al.}(2006)\citenamefont
  {Copeland}, \citenamefont {Pompliano},\ and\ \citenamefont
  {Meek}}]{copeland06}%
  \BibitemOpen
  \bibfield  {author} {\bibinfo {author} {\bibfnamefont {R.~A.}\ \bibnamefont
  {Copeland}}, \bibinfo {author} {\bibfnamefont {D.~L.}\ \bibnamefont
  {Pompliano}}, \ and\ \bibinfo {author} {\bibfnamefont {T.~D.}\ \bibnamefont
  {Meek}},\ }\href@noop {} {\bibfield  {journal} {\bibinfo  {journal} {Nature
  Reviews: Drug Discovery}\ }\textbf {\bibinfo {volume} {5}},\ \bibinfo {pages}
  {731} (\bibinfo {year} {2006})}\BibitemShut {NoStop}%
\bibitem [{\citenamefont {Berne}\ \emph {et~al.}(2009)\citenamefont {Berne},
  \citenamefont {Weeks},\ and\ \citenamefont {Zhou}}]{berne_rev}%
  \BibitemOpen
  \bibfield  {author} {\bibinfo {author} {\bibfnamefont {B.~J.}\ \bibnamefont
  {Berne}}, \bibinfo {author} {\bibfnamefont {J.}~\bibnamefont {Weeks}}, \ and\
  \bibinfo {author} {\bibfnamefont {R.}~\bibnamefont {Zhou}},\ }\href@noop {}
  {\bibfield  {journal} {\bibinfo  {journal} {{Ann. Rev. Phys. Chem. }}\
  }\textbf {\bibinfo {volume} {60}},\ \bibinfo {pages} {85} (\bibinfo {year}
  {2009})}\BibitemShut {NoStop}%
\bibitem [{\citenamefont {Rasaiah}\ \emph {et~al.}(2008)\citenamefont
  {Rasaiah}, \citenamefont {Garde},\ and\ \citenamefont {Hummer}}]{Hummer_rev}%
  \BibitemOpen
  \bibfield  {author} {\bibinfo {author} {\bibfnamefont {J.~C.}\ \bibnamefont
  {Rasaiah}}, \bibinfo {author} {\bibfnamefont {S.}~\bibnamefont {Garde}}, \
  and\ \bibinfo {author} {\bibfnamefont {G.}~\bibnamefont {Hummer}},\
  }\href@noop {} {\bibfield  {journal} {\bibinfo  {journal} {{Ann. Rev. Phys.
  Chem. }}\ }\textbf {\bibinfo {volume} {59}},\ \bibinfo {pages} {713}
  (\bibinfo {year} {2008})}\BibitemShut {NoStop}%
\bibitem [{\citenamefont {Chandler}(2005)}]{chandler_rev}%
  \BibitemOpen
  \bibfield  {author} {\bibinfo {author} {\bibfnamefont {D.}~\bibnamefont
  {Chandler}},\ }\href@noop {} {\bibfield  {journal} {\bibinfo  {journal}
  {Nature}\ }\textbf {\bibinfo {volume} {437}},\ \bibinfo {pages} {640}
  (\bibinfo {year} {2005})}\BibitemShut {NoStop}%
\bibitem [{\citenamefont {Sharma}\ and\ \citenamefont
  {Debenedetti}(2012{\natexlab{a}})}]{sharma12}%
  \BibitemOpen
  \bibfield  {author} {\bibinfo {author} {\bibfnamefont {S.}~\bibnamefont
  {Sharma}}\ and\ \bibinfo {author} {\bibfnamefont {P.~G.}\ \bibnamefont
  {Debenedetti}},\ }\href@noop {} {\bibfield  {journal} {\bibinfo  {journal}
  {Proc. Natl. Acad. Sci. USA}\ }\textbf {\bibinfo {volume} {109}},\ \bibinfo
  {pages} {4365} (\bibinfo {year} {2012}{\natexlab{a}})}\BibitemShut {NoStop}%
\bibitem [{\citenamefont {Sharma}\ and\ \citenamefont
  {Debenedetti}(2012{\natexlab{b}})}]{sharma12-2}%
  \BibitemOpen
  \bibfield  {author} {\bibinfo {author} {\bibfnamefont {S.}~\bibnamefont
  {Sharma}}\ and\ \bibinfo {author} {\bibfnamefont {P.~G.}\ \bibnamefont
  {Debenedetti}},\ }\href@noop {} {\bibfield  {journal} {\bibinfo  {journal}
  {J. Phys. Chem. B}\ }\textbf {\bibinfo {volume} {116}},\ \bibinfo {pages}
  {13282} (\bibinfo {year} {2012}{\natexlab{b}})}\BibitemShut {NoStop}%
\bibitem [{\citenamefont {Liu}\ \emph {et~al.}(2005)\citenamefont {Liu},
  \citenamefont {Huang}, \citenamefont {Zhou},\ and\ \citenamefont
  {Berne}}]{liu05}%
  \BibitemOpen
  \bibfield  {author} {\bibinfo {author} {\bibfnamefont {P.}~\bibnamefont
  {Liu}}, \bibinfo {author} {\bibfnamefont {X.}~\bibnamefont {Huang}}, \bibinfo
  {author} {\bibfnamefont {R.}~\bibnamefont {Zhou}}, \ and\ \bibinfo {author}
  {\bibfnamefont {B.~J.}\ \bibnamefont {Berne}},\ }\href@noop {} {\bibfield
  {journal} {\bibinfo  {journal} {Nature}\ }\textbf {\bibinfo {volume} {437}},\
  \bibinfo {pages} {159} (\bibinfo {year} {2005})}\BibitemShut {NoStop}%
\bibitem [{\citenamefont {Hummer}\ \emph {et~al.}(2001)\citenamefont {Hummer},
  \citenamefont {Rasaiah},\ and\ \citenamefont {Noworyta}}]{Hummer_cnt}%
  \BibitemOpen
  \bibfield  {author} {\bibinfo {author} {\bibfnamefont {G.}~\bibnamefont
  {Hummer}}, \bibinfo {author} {\bibfnamefont {J.~C.}\ \bibnamefont {Rasaiah}},
  \ and\ \bibinfo {author} {\bibfnamefont {J.}~\bibnamefont {Noworyta}},\
  }\href@noop {} {\bibfield  {journal} {\bibinfo  {journal} {{Nature }}\
  }\textbf {\bibinfo {volume} {414}},\ \bibinfo {pages} {188} (\bibinfo {year}
  {2001})}\BibitemShut {NoStop}%
\bibitem [{\citenamefont {Morrone}\ \emph {et~al.}(2012)\citenamefont
  {Morrone}, \citenamefont {Li},\ and\ \citenamefont {Berne}}]{joe12}%
  \BibitemOpen
  \bibfield  {author} {\bibinfo {author} {\bibfnamefont {J.~A.}\ \bibnamefont
  {Morrone}}, \bibinfo {author} {\bibfnamefont {J.}~\bibnamefont {Li}}, \ and\
  \bibinfo {author} {\bibfnamefont {B.~J.}\ \bibnamefont {Berne}},\ }\href@noop
  {} {\bibfield  {journal} {\bibinfo  {journal} {J. Phys. Chem. B}\ }\textbf
  {\bibinfo {volume} {116}},\ \bibinfo {pages} {378} (\bibinfo {year}
  {2012})}\BibitemShut {NoStop}%
\bibitem [{\citenamefont {Li}\ \emph {et~al.}(2012)\citenamefont {Li},
  \citenamefont {Morrone},\ and\ \citenamefont {Berne}}]{jingyuan12}%
  \BibitemOpen
  \bibfield  {author} {\bibinfo {author} {\bibfnamefont {J.}~\bibnamefont
  {Li}}, \bibinfo {author} {\bibfnamefont {J.~A.}\ \bibnamefont {Morrone}}, \
  and\ \bibinfo {author} {\bibfnamefont {B.~J.}\ \bibnamefont {Berne}},\
  }\href@noop {} {\bibfield  {journal} {\bibinfo  {journal} {J. Phys. Chem. B}\
  }\textbf {\bibinfo {volume} {116}},\ \bibinfo {pages} {11537} (\bibinfo
  {year} {2012})}\BibitemShut {NoStop}%
\bibitem [{\citenamefont {Baron}\ \emph {et~al.}(2010)\citenamefont {Baron},
  \citenamefont {Setny},\ and\ \citenamefont {McCammon}}]{Baron:JACS10}%
  \BibitemOpen
  \bibfield  {author} {\bibinfo {author} {\bibfnamefont {R.}~\bibnamefont
  {Baron}}, \bibinfo {author} {\bibfnamefont {P.}~\bibnamefont {Setny}}, \ and\
  \bibinfo {author} {\bibfnamefont {J.~A.}\ \bibnamefont {McCammon}},\
  }\href@noop {} {\bibfield  {journal} {\bibinfo  {journal} {J. Am. Chem.
  Soc.}\ }\textbf {\bibinfo {volume} {132}},\ \bibinfo {pages} {12091}
  (\bibinfo {year} {2010})}\BibitemShut {NoStop}%
\bibitem [{\citenamefont {Baron}\ and\ \citenamefont
  {McCammon}(2012)}]{baron13}%
  \BibitemOpen
  \bibfield  {author} {\bibinfo {author} {\bibfnamefont {R.}~\bibnamefont
  {Baron}}\ and\ \bibinfo {author} {\bibfnamefont {J.~A.}\ \bibnamefont
  {McCammon}},\ }\href@noop {} {\bibfield  {journal} {\bibinfo  {journal}
  {Annu. Rev. Phys. Chem.}\ }\textbf {\bibinfo {volume} {64}},\ \bibinfo
  {pages} {151} (\bibinfo {year} {2012})}\BibitemShut {NoStop}%
\bibitem [{\citenamefont {Tully}(2012)}]{tully12}%
  \BibitemOpen
  \bibfield  {author} {\bibinfo {author} {\bibfnamefont {J.~C.}\ \bibnamefont
  {Tully}},\ }\href@noop {} {\bibfield  {journal} {\bibinfo  {journal} {J.
  Chem. Phys.}\ }\textbf {\bibinfo {volume} {137}},\ \bibinfo {pages} {22A301}
  (\bibinfo {year} {2012})}\BibitemShut {NoStop}%
\bibitem [{\citenamefont {Setny}\ \emph {et~al.}(2013)\citenamefont {Setny},
  \citenamefont {Baron}, \citenamefont {Kekenes-Huskey}, \citenamefont
  {McCammon},\ and\ \citenamefont {Dzubiella}}]{setny13}%
  \BibitemOpen
  \bibfield  {author} {\bibinfo {author} {\bibfnamefont {P.}~\bibnamefont
  {Setny}}, \bibinfo {author} {\bibfnamefont {R.}~\bibnamefont {Baron}},
  \bibinfo {author} {\bibfnamefont {P.~M.}\ \bibnamefont {Kekenes-Huskey}},
  \bibinfo {author} {\bibfnamefont {J.~A.}\ \bibnamefont {McCammon}}, \ and\
  \bibinfo {author} {\bibfnamefont {J.}~\bibnamefont {Dzubiella}},\ }\href@noop
  {} {\bibfield  {journal} {\bibinfo  {journal} {Proc. Natl. Acad. Sci. USA}\
  }\textbf {\bibinfo {volume} {110}},\ \bibinfo {pages} {1197} (\bibinfo {year}
  {2013})}\BibitemShut {NoStop}%
\bibitem [{\citenamefont {Willard}\ and\ \citenamefont
  {Chandler}(2008)}]{willard08}%
  \BibitemOpen
  \bibfield  {author} {\bibinfo {author} {\bibfnamefont {A.~P.}\ \bibnamefont
  {Willard}}\ and\ \bibinfo {author} {\bibfnamefont {D.}~\bibnamefont
  {Chandler}},\ }\href@noop {} {\bibfield  {journal} {\bibinfo  {journal} {J.
  Phys. Chem. B}\ }\textbf {\bibinfo {volume} {112}},\ \bibinfo {pages} {6187}
  (\bibinfo {year} {2008})}\BibitemShut {NoStop}%
\bibitem [{\citenamefont {Humphrey}\ \emph {et~al.}(1996)\citenamefont
  {Humphrey}, \citenamefont {Dalke},\ and\ \citenamefont {Schulten}}]{vmd96}%
  \BibitemOpen
  \bibfield  {author} {\bibinfo {author} {\bibfnamefont {W.}~\bibnamefont
  {Humphrey}}, \bibinfo {author} {\bibfnamefont {A.}~\bibnamefont {Dalke}}, \
  and\ \bibinfo {author} {\bibfnamefont {K.}~\bibnamefont {Schulten}},\
  }\href@noop {} {\bibfield  {journal} {\bibinfo  {journal} {J. Molec.
  Graphics}\ }\textbf {\bibinfo {volume} {14}},\ \bibinfo {pages} {33}
  (\bibinfo {year} {1996})}\BibitemShut {NoStop}%
\bibitem [{\citenamefont {Berne}\ and\ \citenamefont
  {Pecora}(1990)}]{bernebook}%
  \BibitemOpen
  \bibfield  {author} {\bibinfo {author} {\bibfnamefont {B.~J.}\ \bibnamefont
  {Berne}}\ and\ \bibinfo {author} {\bibfnamefont {R.}~\bibnamefont {Pecora}},\
  }\href@noop {} {\emph {\bibinfo {title} {Dynamic Light Scattering}}}\ (\bibinfo  {publisher}
  {Dover},\ \bibinfo {year} {1990})\BibitemShut {NoStop}%
\bibitem [{\citenamefont {Bicout}\ and\ \citenamefont {Szabo}(1998)}]{szabo98}%
  \BibitemOpen
  \bibfield  {author} {\bibinfo {author} {\bibfnamefont {D.~J.}\ \bibnamefont
  {Bicout}}\ and\ \bibinfo {author} {\bibfnamefont {A.}~\bibnamefont {Szabo}},\
  }\href@noop {} {\bibfield  {journal} {\bibinfo  {journal} {J. Chem. Phys.}\
  }\textbf {\bibinfo {volume} {109}},\ \bibinfo {pages} {2325} (\bibinfo {year}
  {1998})}\BibitemShut {NoStop}%
\bibitem [{\citenamefont {Pande}\ \emph {et~al.}(2010)\citenamefont {Pande},
  \citenamefont {Beauchamp},\ and\ \citenamefont {Bowman}}]{pande2010}%
  \BibitemOpen
  \bibfield  {author} {\bibinfo {author} {\bibfnamefont {V.~S.}\ \bibnamefont
  {Pande}}, \bibinfo {author} {\bibfnamefont {K.}~\bibnamefont {Beauchamp}}, \
  and\ \bibinfo {author} {\bibfnamefont {G.~R.}\ \bibnamefont {Bowman}},\
  }\href@noop {} {\bibfield  {journal} {\bibinfo  {journal} {Methods}\ }\textbf
  {\bibinfo {volume} {52}},\ \bibinfo {pages} {99} (\bibinfo {year}
  {2010})}\BibitemShut {NoStop}%
\bibitem [{\citenamefont {Sriraman}\ \emph {et~al.}(2005)\citenamefont
  {Sriraman}, \citenamefont {Kevrekidis},\ and\ \citenamefont
  {Hummer}}]{sriraman05}%
  \BibitemOpen
  \bibfield  {author} {\bibinfo {author} {\bibfnamefont {S.}~\bibnamefont
  {Sriraman}}, \bibinfo {author} {\bibfnamefont {I.~G.}\ \bibnamefont
  {Kevrekidis}}, \ and\ \bibinfo {author} {\bibfnamefont {G.}~\bibnamefont
  {Hummer}},\ }\href@noop {} {\bibfield  {journal} {\bibinfo  {journal} {J.
  Phys. Chem. B}\ }\textbf {\bibinfo {volume} {109}},\ \bibinfo {pages} {6479}
  (\bibinfo {year} {2005})}\BibitemShut {NoStop}%
\bibitem [{\citenamefont {Gu}\ \emph {et~al.}(2013)\citenamefont {Gu},
  \citenamefont {Chang}, \citenamefont {Maibaum}, \citenamefont {Pande},
  \citenamefont {Carlsson},\ and\ \citenamefont {Guibas}}]{gu2013}%
  \BibitemOpen
  \bibfield  {author} {\bibinfo {author} {\bibfnamefont {C.}~\bibnamefont
  {Gu}}, \bibinfo {author} {\bibfnamefont {H.}~\bibnamefont {Chang}}, \bibinfo
  {author} {\bibfnamefont {L.}~\bibnamefont {Maibaum}}, \bibinfo {author}
  {\bibfnamefont {V.~S.}\ \bibnamefont {Pande}}, \bibinfo {author}
  {\bibfnamefont {G.~E.}\ \bibnamefont {Carlsson}}, \ and\ \bibinfo {author}
  {\bibfnamefont {L.~J.}\ \bibnamefont {Guibas}},\ }\href@noop {} {\bibfield
  {journal} {\bibinfo  {journal} {BMC Bioinformatics}\ }\textbf {\bibinfo
  {volume} {14(Suppl 2)}},\ \bibinfo {pages} {S8} (\bibinfo {year}
  {2013})}\BibitemShut {NoStop}%
\bibitem [{\citenamefont {Bocquet}\ \emph {et~al.}(1997)\citenamefont
  {Bocquet}, \citenamefont {Hansen},\ and\ \citenamefont
  {Piasecki}}]{Bocquet:1997p103}%
  \BibitemOpen
  \bibfield  {author} {\bibinfo {author} {\bibfnamefont {L.}~\bibnamefont
  {Bocquet}}, \bibinfo {author} {\bibfnamefont {J.}~\bibnamefont {Hansen}}, \
  and\ \bibinfo {author} {\bibfnamefont {J.}~\bibnamefont {Piasecki}},\
  }\href@noop {} {\bibfield  {journal} {\bibinfo  {journal} {J. Stat. Phys.}\
  }\textbf {\bibinfo {volume} {89}},\ \bibinfo {pages} {321} (\bibinfo {year}
  {1997})}\BibitemShut {NoStop}%
\bibitem [{\citenamefont {Straub}\ \emph {et~al.}(1990)\citenamefont {Straub},
  \citenamefont {Berne},\ and\ \citenamefont {Roux}}]{Straub:1990p65}%
  \BibitemOpen
  \bibfield  {author} {\bibinfo {author} {\bibfnamefont {J.~E.}\ \bibnamefont
  {Straub}}, \bibinfo {author} {\bibfnamefont {B.~J.}\ \bibnamefont {Berne}}, \
  and\ \bibinfo {author} {\bibfnamefont {B.}~\bibnamefont {Roux}},\ }\href@noop
  {} {\bibfield  {journal} {\bibinfo  {journal} {J. Chem. Phys.}\ }\textbf
  {\bibinfo {volume} {93}},\ \bibinfo {pages} {6804} (\bibinfo {year}
  {1990})}\BibitemShut {NoStop}%
\bibitem [{\citenamefont {Hummer}(2005)}]{hummer05}%
  \BibitemOpen
  \bibfield  {author} {\bibinfo {author} {\bibfnamefont {G.}~\bibnamefont
  {Hummer}},\ }\href@noop {} {\bibfield  {journal} {\bibinfo  {journal} {New J.
  Phys.}\ }\textbf {\bibinfo {volume} {7}},\ \bibinfo {pages} {34} (\bibinfo
  {year} {2005})}\BibitemShut {NoStop}%
\bibitem [{\citenamefont {Giovambattista}\ \emph {et~al.}(2007)\citenamefont
  {Giovambattista}, \citenamefont {Debenedetti},\ and\ \citenamefont
  {Rossky}}]{rossky07}%
  \BibitemOpen
  \bibfield  {author} {\bibinfo {author} {\bibfnamefont {N.}~\bibnamefont
  {Giovambattista}}, \bibinfo {author} {\bibfnamefont {P.~G.}\ \bibnamefont
  {Debenedetti}}, \ and\ \bibinfo {author} {\bibfnamefont {P.~J.}\ \bibnamefont
  {Rossky}},\ }\href@noop {} {\bibfield  {journal} {\bibinfo  {journal} {J.
  Phys. Chem. C}\ }\textbf {\bibinfo {volume} {111}},\ \bibinfo {pages} {1323}
  (\bibinfo {year} {2007})}\BibitemShut {NoStop}%
\bibitem [{\citenamefont {Jorgensen}\ \emph {et~al.}(1983)\citenamefont
  {Jorgensen}, \citenamefont {Chandrasekhar}, \citenamefont {Madura},
  \citenamefont {Impey},\ and\ \citenamefont {Klein}}]{tip4p}%
  \BibitemOpen
  \bibfield  {author} {\bibinfo {author} {\bibfnamefont {W.}~\bibnamefont
  {Jorgensen}}, \bibinfo {author} {\bibfnamefont {J.}~\bibnamefont
  {Chandrasekhar}}, \bibinfo {author} {\bibfnamefont {J.}~\bibnamefont
  {Madura}}, \bibinfo {author} {\bibfnamefont {R.}~\bibnamefont {Impey}}, \
  and\ \bibinfo {author} {\bibfnamefont {M.}~\bibnamefont {Klein}},\
  }\href@noop {} {\bibfield  {journal} {\bibinfo  {journal} {{J.Chem.Phys. }}\
  }\textbf {\bibinfo {volume} {79}},\ \bibinfo {pages} {926} (\bibinfo {year}
  {1983})}\BibitemShut {NoStop}%
\bibitem [{\citenamefont {Bussi}\ \emph {et~al.}(2007)\citenamefont {Bussi},
  \citenamefont {Donadio},\ and\ \citenamefont {Parrinello}}]{Bussi:2007p114}%
  \BibitemOpen
  \bibfield  {author} {\bibinfo {author} {\bibfnamefont {G.}~\bibnamefont
  {Bussi}}, \bibinfo {author} {\bibfnamefont {D.}~\bibnamefont {Donadio}}, \
  and\ \bibinfo {author} {\bibfnamefont {M.}~\bibnamefont {Parrinello}},\
  }\href@noop {} {\bibfield  {journal} {\bibinfo  {journal} {J. Chem. Phys.}\
  }\textbf {\bibinfo {volume} {126}},\ \bibinfo {pages} {014101} (\bibinfo
  {year} {2007})}\BibitemShut {NoStop}%
\bibitem [{\citenamefont {Berendsen}\ \emph {et~al.}(1984)\citenamefont
  {Berendsen}, \citenamefont {Postma}, \citenamefont {van Gunsteren},
  \citenamefont {DiNola},\ and\ \citenamefont {Haak}}]{berendsen}%
  \BibitemOpen
  \bibfield  {author} {\bibinfo {author} {\bibfnamefont {H.~J.~C.}\
  \bibnamefont {Berendsen}}, \bibinfo {author} {\bibfnamefont {J.~P.~M.}\
  \bibnamefont {Postma}}, \bibinfo {author} {\bibfnamefont {W.~F.}\
  \bibnamefont {van Gunsteren}}, \bibinfo {author} {\bibfnamefont
  {A.}~\bibnamefont {DiNola}}, \ and\ \bibinfo {author} {\bibfnamefont {J.~R.}\
  \bibnamefont {Haak}},\ }\href@noop {} {\bibfield  {journal} {\bibinfo
  {journal} {J. Chem. Phys.}\ }\textbf {\bibinfo {volume} {81}},\ \bibinfo
  {pages} {3684} (\bibinfo {year} {1984})}\BibitemShut {NoStop}%
\bibitem [{\citenamefont {Hess}\ \emph {et~al.}(2008)\citenamefont {Hess},
  \citenamefont {Kutzner},\ and\ \citenamefont {van~der Spoel}}]{gromacs4}%
  \BibitemOpen
  \bibfield  {author} {\bibinfo {author} {\bibfnamefont {B.}~\bibnamefont
  {Hess}}, \bibinfo {author} {\bibfnamefont {C.}~\bibnamefont {Kutzner}}, \
  and\ \bibinfo {author} {\bibfnamefont {D.}~\bibnamefont {van~der Spoel}},\
  }\href@noop {} {\bibfield  {journal} {\bibinfo  {journal} {J. Chem. Theor.
  Comput.}\ }\textbf {\bibinfo {volume} {4}},\ \bibinfo {pages} {435} (\bibinfo
  {year} {2008})}\BibitemShut {NoStop}%
\bibitem [{\citenamefont {Bonomi}\ \emph {et~al.}(2009)\citenamefont {Bonomi},
  \citenamefont {Branduardi}, \citenamefont {Bussi}, \citenamefont {Camilloni},
  \citenamefont {Provasi}, \citenamefont {Raiteri}, \citenamefont {Donadio},
  \citenamefont {Marinelli}, \citenamefont {Pietrucci}, \citenamefont
  {Broglia},\ and\ \citenamefont {Parrinello}}]{plumed}%
  \BibitemOpen
  \bibfield  {author} {\bibinfo {author} {\bibfnamefont {M.}~\bibnamefont
  {Bonomi}}, \bibinfo {author} {\bibfnamefont {D.}~\bibnamefont {Branduardi}},
  \bibinfo {author} {\bibfnamefont {G.}~\bibnamefont {Bussi}}, \bibinfo
  {author} {\bibfnamefont {C.}~\bibnamefont {Camilloni}}, \bibinfo {author}
  {\bibfnamefont {D.}~\bibnamefont {Provasi}}, \bibinfo {author} {\bibfnamefont
  {P.}~\bibnamefont {Raiteri}}, \bibinfo {author} {\bibfnamefont
  {D.}~\bibnamefont {Donadio}}, \bibinfo {author} {\bibfnamefont
  {F.}~\bibnamefont {Marinelli}}, \bibinfo {author} {\bibfnamefont
  {F.}~\bibnamefont {Pietrucci}}, \bibinfo {author} {\bibfnamefont {R.~A.}\
  \bibnamefont {Broglia}}, \ and\ \bibinfo {author} {\bibfnamefont
  {M.}~\bibnamefont {Parrinello}},\ }\href@noop {} {\bibfield  {journal}
  {\bibinfo  {journal} {Comp. Phys. Comm.}\ }\textbf {\bibinfo {volume}
  {180}},\ \bibinfo {pages} {1961} (\bibinfo {year} {2009})}\BibitemShut
  {NoStop}%
\end{thebibliography}
\end{document}